\documentclass[sigconf, dvipsnames]{acmart}
\AtBeginDocument{%
  \providecommand\BibTeX{{%
    \normalfont B\kern-0.5em{\scshape i\kern-0.25em b}\kern-0.8em\TeX}}}

\copyrightyear{2020}
\acmYear{2020}
\setcopyright{acmcopyright}
\acmConference[KDD '20] {26th ACM SIGKDD Conference on Knowledge Discovery and Data Mining}{August 23--27, 2020}{Virtual Event, USA}
\acmBooktitle{26th ACM SIGKDD Conference on Knowledge Discovery and Data Mining (KDD '20), August 23--27, 2020, Virtual Event, USA}
\acmPrice{15.00}
\acmDOI{10.1145/XXXXXX.XXXXXX}
\acmISBN{978-1-4503-7998-4/20/08}

\PassOptionsToPackage{usenames,dvipsnames}{xcolor}

\usepackage{amsmath,amssymb,amsthm}
\usepackage{algorithm}
\usepackage{graphicx}
\usepackage{subcaption}
\usepackage[belowskip=1pt,aboveskip=1pt]{caption}
\usepackage{textcomp}

\usepackage{url}
\usepackage{booktabs}
\usepackage{epstopdf}
\usepackage{multirow}
\usepackage{array}
\usepackage{enumitem}
\usepackage{wrapfig}
\usepackage[many]{tcolorbox}
\usepackage{mathtools}
\usepackage{dsfont}
\usepackage{esvect}
\usepackage{hyperref}
\usepackage{multicol}
\usepackage{xpatch}
\usepackage[noend]{algpseudocode}
\usepackage{comment}

\makeatletter
\xpatchcmd{\algorithmic}{\itemsep\z@}{\itemsep=0.1ex plus0.5pt}{}{}
\makeatother

\DeclareMathOperator{\RI}{RI}
\DeclareMathOperator{\ERI}{ERI}

\DeclareMathOperator{\PE}{PE}

\begin{document}


\title{Robust Spammer Detection by Nash Reinforcement Learning}



\author{Yingtong Dou}
\affiliation{%
  \institution{Univ. of Illinois at Chicago}}
\email{ydou5@uic.edu}

\author{Guixiang Ma$^\ast$}
\affiliation{%
  \institution{Intel Labs}}
\email{guixiang.ma@intel.com}

\author{Philip S. Yu}
\affiliation{%
  \institution{Univ. of Illinois at Chicago}}
\email{psyu@uic.edu}

\author{Sihong Xie}
\affiliation{%
  \institution{Lehigh University}}
\email{xiesihong1@gmail.com}


\begin{abstract}
Online reviews provide product evaluations
for customers to make decisions.
Unfortunately, the evaluations can be manipulated using fake reviews (``spams'') by professional spammers,
who have learned increasingly insidious and powerful spamming strategies by adapting to the deployed detectors.
Spamming strategies are hard to capture, as they can be varying quickly along time, different across spammers and target products, and more critically, remained unknown in most cases.
Furthermore, most existing detectors focus on detection accuracy,
which is not well-aligned
with the goal of maintaining
the trustworthiness of product evaluations.
To address the challenges,
we formulate a minimax game where the spammers and spam detectors compete with each other on their practical goals
that are not solely based on detection accuracy.
Nash equilibria of the game lead to stable detectors that are agnostic to any mixed detection strategies.
However, the game has no closed-form solution
and is not differentiable to admit the typical gradient-based algorithms.
We turn the game into two dependent Markov Decision Processes (MDPs) to allow efficient stochastic optimization based on multi-armed bandit and policy gradient. 
We experiment on three large review datasets using various state-of-the-art spamming and detection strategies and
show that the optimization algorithm can reliably find an equilibrial detector that can robustly 
and effectively prevent spammers with any mixed spamming strategies from attaining their practical goal. Our code is available at \url{https://github.com/YingtongDou/Nash-Detect}.
\end{abstract}

\maketitle

\section{Introduction}

Online reviews and ratings contributed by real customers
help shape reputations of the businesses and guide customer decision-makings, playing an integrative role in e-commerce and websites such as Amazon~\cite{forman2008examining}, Yelp~\cite{luca2016reviews}, and Google Play~\cite{rahman2019art}. However, monetary incentives therein have also attracted a large number of spammers to hold sway over less informed customers:
it is estimated that about 40\% of the reviews on Amazon are fake (called ``review spams'')~\cite{nymag}.
To cope with the spams and restore the trustworthiness of online reviews,
many detection methods based on texts~\cite{Mukherjee:2013uk, wang2017handling, kaghazgaran2019wide}, reviewer behaviors~\cite{kumar2018rev2, Xie2012a,mukherjee2013spotting}, and graphs~\cite{Rayana2015,Hooi2016, kaghazgaran2018combating, liu2020alleviating} have been proposed. See Table~\ref{tab:compare_work} for some state-of-the-art.

\begin{table}[h]
\centering
\caption{\footnotesize Comparison of the proposed Nash-Detect and prior work.
``Attack Agnostic'' indicates a method does not assume a fixed attack.
``Practical Goal'' means at a business metric is considered.
}
\resizebox{\linewidth}{!}{%
\begin{tabular}{r|cc|ccccc|c}
\multirow{2}{*} &  \multicolumn{2}{c|}{\textbf{Business}} & 
\multicolumn{5}{c|}{\textbf{Fraud Detection}} & \multirow{2}{*}{\textbf{\shortstack{Proposed \\ Nash-Detect} }} \\
\cline{2-8}
  & \cite{lappas2016impact} & \cite{luca2016fake} & \cite{zheng2017smoke} & \cite{Hooi2016} & \cite{Yao:2017bd} &  \cite{kaghazgaran2019tomcat}&  \cite{breuer2020friend} &\\\hline
Practical Goal  & $\checkmark$ & $\checkmark$ & $\checkmark$ & & $\checkmark$ & $\checkmark$ & &$\checkmark$\\
 Attack Agnostic   &  &  & & $\checkmark$ & & & $\checkmark$ & $\checkmark$\\

\end{tabular}}
\label{tab:compare_work}
\end{table}

\begin{figure*}
\centering
\begin{subfigure}[b]{0.7\textwidth}
   \includegraphics[width=\linewidth]{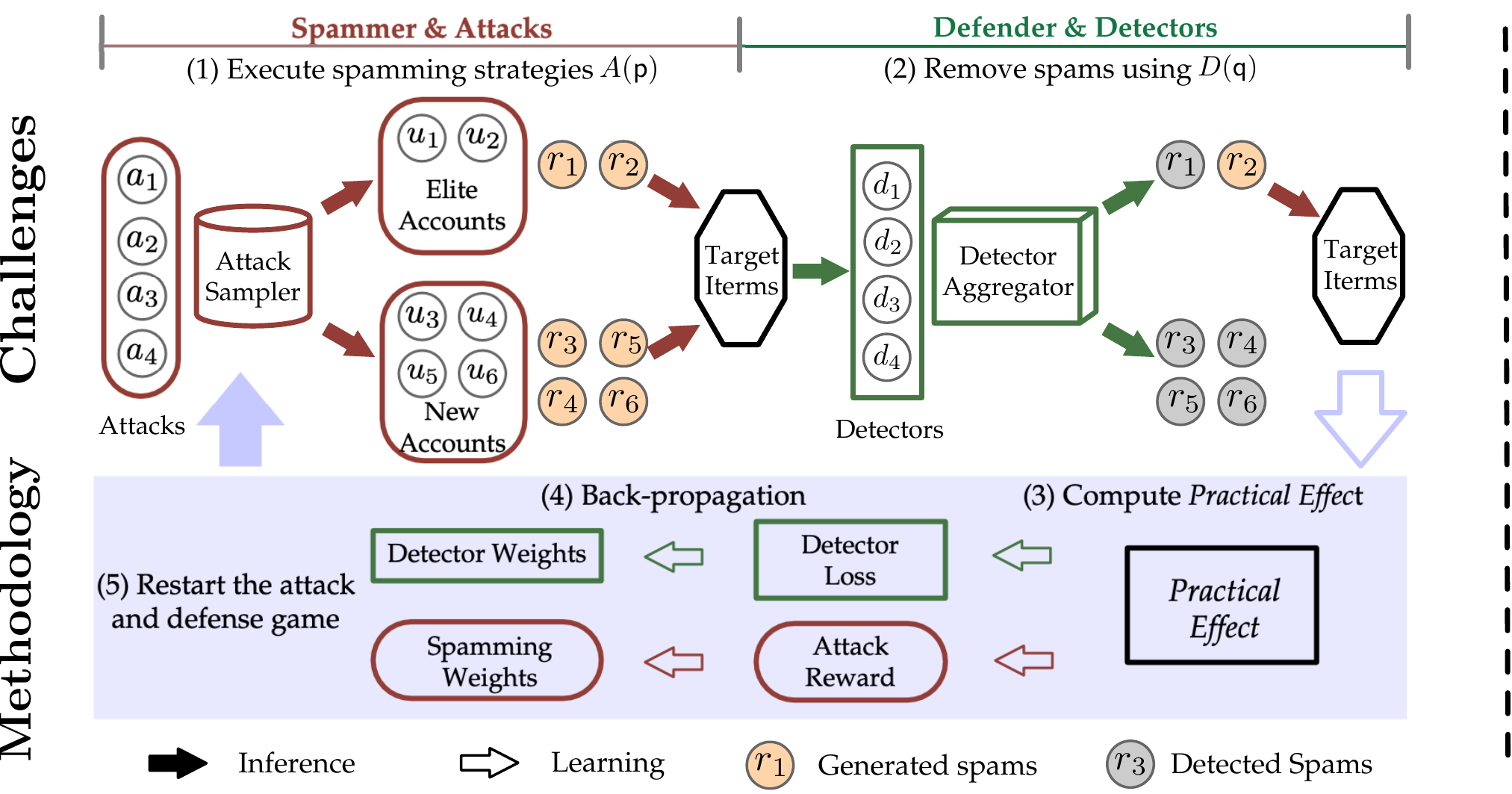}
   \caption{\footnotesize Challenges and our methodology.}
   \label{fig:motivations}
\end{subfigure}
\hfill
\begin{subfigure}[b]{0.295\textwidth}
\centering
\begin{subfigure}[b]{0.78\textwidth}
\centering
   \includegraphics[width=\linewidth]{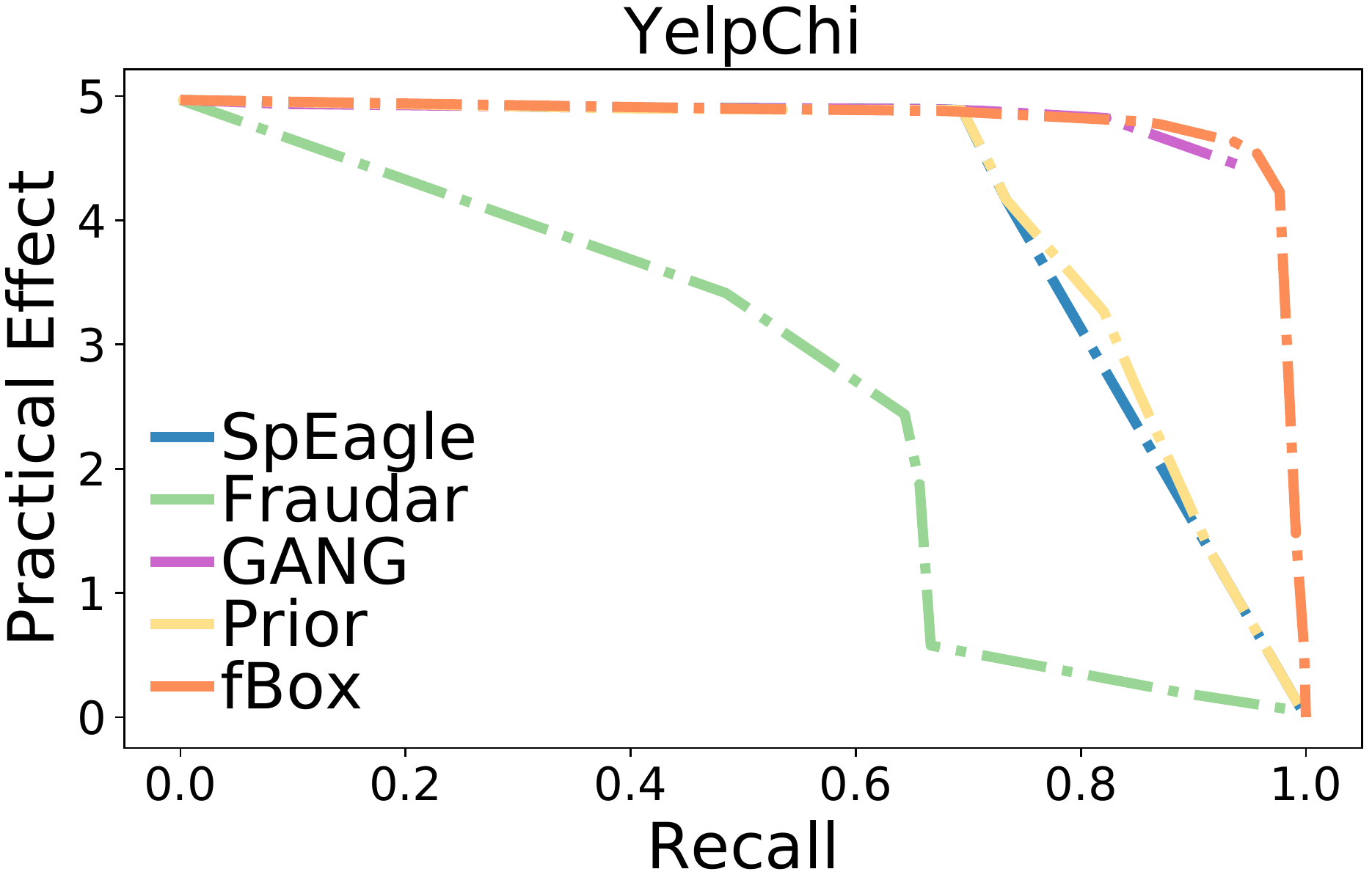}
   \caption{\footnotesize Practical Effect vs. Recall for different detectors.}
   \label{fig:pe_f1}
\end{subfigure}

\bigskip
\begin{subfigure}[b]{0.78\textwidth}
\centering
   \includegraphics[width=\linewidth]{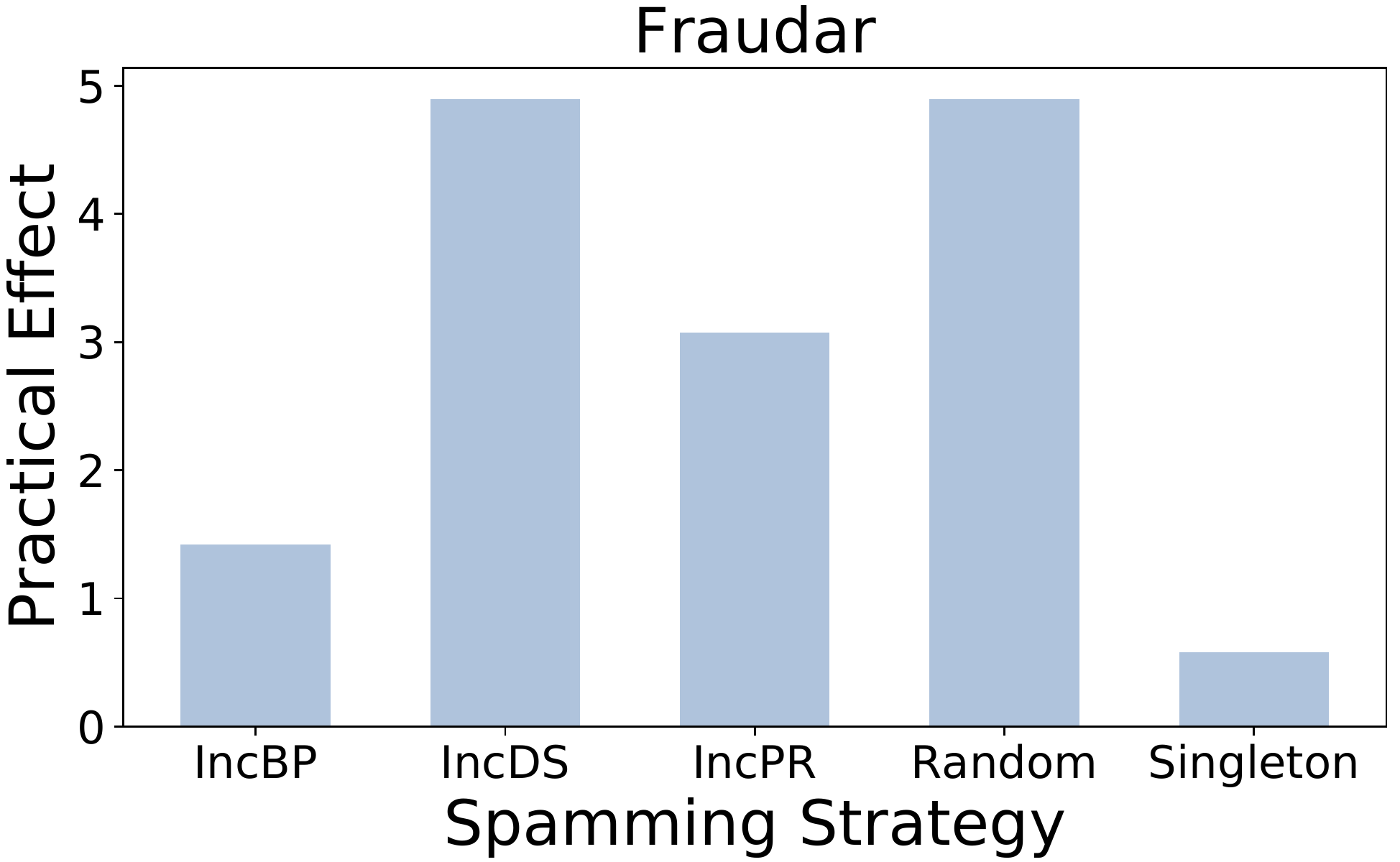}
   \caption{\footnotesize Practical effect under different spamming attack strategies.}
   \label{fig:diff_pe}
\end{subfigure}
\end{subfigure}
\caption{\footnotesize \textcolor{Periwinkle}{(a):} A vulnerable spam detection pipeline, with steps numbered as (1), (2), etc. Accuracy-based detectors can be misled to detect numerous insignificant spams from new accounts,
leaving behind the more manipulative elite spams.
We define a zero-sum game to find a robust defender against unknown and evolving spamming strategies $A(p)$. \textcolor{Periwinkle}{(b):} The Practical Effect vs. Recall of individual detectors (shown in legend) against a mixed spamming strategy. The curve is obtained by sweeping the detection thresholds.
For most detectors, the attack could attain high practical effects even with high detection recall scores. \textcolor{Periwinkle}{(c):} For a fixed spam detector (Fraudar), a spammer can choose the best out of five attack strategies to maximize the practical effect.}
\end{figure*}

We note two drawbacks of existing detectors.
\textbf{1)} Most detectors assume spams are generated by spammers\footnote{A ``spammer'' refers to a physical person or entity in the real world that spams, rather than a ``spamming account'' that is set up by a spammer on the online review system.\\
$\ast$This work was done when the author was at University of Illinois at Chicago.}
with the same mindset and can detect the spams by relying on the assumed spamming strategy.
In the real world, there are multiple groups of spammers who have different goals, targets, resources, and strategies.
One spammer may want to promote a new business while another aims to demote a popular brand's competitors~\cite{luca2016fake}.
The wide spectrum of detection signals published so far
serves as strong evidence that multiple spamming strategies co-exist and no single detector can stop the spams.
\textbf{2)}
Professional spammers are more persistent and committed, and can research the latest detection techniques from published papers~\cite{Mukherjee:2013uk}, third-party detection websites with detailed detection manuals~\cite{reviewmeta}, and labeled spams released by the review website~\cite{Rayana2015}.
The spammer can integrate these resources and learn to infiltrate the deployed detectors.

The above drawbacks are partially addressed by a diverse set of spamming modes considered in the prior work~\cite{Hooi2016, Chen2017, yang2020secure, kaghazgaran2019tomcat, wang2019, breuer2020friend}.
Nonetheless,
the prior work aimed at high detection recall, AUC, or top-$k$ accuracy.
Without considering practical spamming goals,
the accuracy-oriented detectors may not directly counteract the unknown spamming goal.
For example, as shown in Figure~\ref{fig:motivations},
when the human inspectors can only screen the most suspicious reviews identified by the detector,
one can achieve high top-$k$ accuracy by reporting $k$ easy-to-catch but insignificant spams,
while letting go the more significant ones that are actually manipulating the review system~\cite{zhou2014cost}.
In other words, the accuracy-oriented detectors and the limited human screening resources create a unique vulnerability in the spam detection pipeline.

We reconsider spam detection by posing the following questions.
\textbf{1)} What is the ultimate intention of the spammers?
What is considered as a successful spamming campaign?
From a marketing perspective,
a spammer aims to maximize reputation manipulations and get paid by dishonest businesses.
We distinguish the manipulative effect by elite accounts from that by regular accounts,
and adopt marketing research results~\cite{luca2016reviews, lappas2016impact}
to formulate a practical quantitative spamming goal.
To show the vulnerability mentioned above, we create spamming attacks against various state-of-the-art spam detectors.
The attacks effectively attain the practical spamming goals while saturating the detectors to rather high detection recall scores (Figure~\ref{fig:pe_f1}).
\textbf{2)} What if there are multiple evolving spamming strategies?
Our experiment results in Figure~\ref{fig:diff_pe}
show that a single fixed detector will always fail to detect some spams when a spammer has knowledge about the detector to carefully craft evasive spams.
This implies a black-box attack: if a fixed detector is adopted,
a spammer can probe the detection system by continuously changing the spam generation strategy until success.
Further, if multiple spamming strategies co-exist, either because a spammer diversifies its strategy portfolio, or because multiple spammers are adopting different strategies simultaneously,
likely, some strategies will successfully infiltrate the fixed detector.

To answer the above questions,
we propose Nash-Detect, an attack-agnostic spam detector that is a Nash equilibrium of a minimax game and can tame adaptive and evolving spamming.
\textbf{1)} We define practical goals for both spammers and detectors from the business perspective. We calibrate the detector via a cost-sensitive loss function focusing on practical spamming effect measured by revenue manipulations rather than detection accuracy.
By minimizing this loss function, the detector can be trained to go after a small number of more manipulative spams, while letting go spams that are less significant.
\textbf{2)} To reveal the vulnerability, we design strong \textit{base} spamming strategies against specific state-of-the-art detectors.
The base strategies will be combined into \textit{mixed} strategies to explore even stronger attacks.
Compared to neural network-based attacks~\cite{zugner2018adversarial,sun2018adversarial},
our spamming strategies rely on unlabeled data and have no parameters 
to train.
A spamming strategy based on parameter learning on a labeled set will be less general and can overfit the training data.
\textbf{3)} We formulate a minimax game, where the spammer will vary its mixed spamming strategies to maximize the practical spamming effect, while the detector will reconfigure its detection strategy to minimize the updated practical spamming effect.
The two strategies will hopefully evolve to an equilibrium consisting of a robust detection strategy.
To ensure computation tractability, we propose a reinforcement learning approach to find an equilibrium of the minimax game in multiple episodes of fictitious plays of the game.
Each episode has two phases.
In the inference phase, the spammer samples a base spamming strategy according to the current mixed strategy, and the detector runs its current detection strategy.
This step will evaluate the practical spamming effect under the current two strategies.
In the learning phase,
we update the two strategies by backpropagating the practical effect to the strategy parameters, so that both players can move in opposite directions to maximize and minimize the effect.
After multiple rounds of
fictitious play, the converged detection strategy will be robust to any spamming strategy, including the worst one that the spammer can synthesize using the base spamming strategies.

Experiments show that Nash-Detect can find the best detector configurations which always have better defending performance than the worst cases the spammer can synthesize.
Nash-Detect also exhibits great stability under various settings during training. 
Experiment results on deployment suggest that one should incorporate diverse accounts and products during training to guarantee the robustness of the resulting detectors.

\section{Preliminaries and Challenges}
\label{sec:prelim}
The data of a review system can be represented by a bipartite graph $\mathcal{G}$ consisting of a set of accounts
$\mathcal{U}=\{u_1,\dots, u_m\}$, a set of products $\mathcal{V}=\{v_1,\dots, v_n\}$ where $m$ is the number of accounts and $n$ is the number of products,
and a set of edges
$\mathcal{R}=\{r_{ij}: i\in \{1,\dots, n\}, j\in \{1,\dots, m\}\}$.
Account $u_i$ posts a review to product $v_j$ if and only if there is $r_{ij}\in\mathcal{R}$ and we use $\mathcal{R}$ and $r_{ij}$ to denote the set of reviews and a specific review.
Node $v_i$ ($u_j$, edge $r_{ij}$, resp.) can have various attributes,
represented by the vector $\mathbf{x}_i$ ($\mathbf{x}_j$, $\mathbf{x}_{ij}$, resp.),
describing account and product profiles, and
review metadata (e.g., posting time, ratings, and the related texts or image contents).
The review system evaluates a node $v\in\mathcal{V}$ with a reputation score $s(v)$, 
according to the product ratings, rankings, and credibility of the review account .
Users may rely on $s(v)$ to decide whether to commit trust, attention, and money to $v$.
For example, elite accounts have higher credibility and more likely to be trusted by users~\cite{luca2016fake, forman2008examining}.

A physical spammer is a person who registers and controls a set of accounts (the ``Sybils'')
$\mathcal{U}_{S}=\mathcal{U}_{E}\cup \mathcal{U}_{N}\subset \mathcal{U}$~\cite{zheng2017smoke},
where $\mathcal{U}_E$ is a set of elite accounts carrying more weights in their ratings,
and $\mathcal{U}_N$ is a set of new accounts that the spammer register at any time. 
When committing a spamming campaign,
the spammer creates a set of spams, represented as a set of new edges $\mathcal{R}_{S}$
emitting from some accounts in $\mathcal{U}_{S}$ towards the target nodes $\mathcal{V}_{T}\subset \mathcal{V}$
and some non-targets $\mathcal{V}\setminus \mathcal{V}_T$ (for camouflaging) within a time window.
For simplicity, we assume the spams are posted towards the target nodes only.
The edges $\mathcal{R}_{S}$ and the attributes of the edges
are decided by a spamming strategy with the goal of increasing or decreasing $s(v)$ for $v\in \mathcal{V}_T$.
A spammer can adopt multiple spamming strategies simultaneously. We assume there are $K$ \textit{base} attack strategies $\mathsf{a}=[a_1,\dots, a_K]$ (specified in Section~\ref{sec:attacks_detectors}).
Given the current review system status and a spamming goal, a spamming strategy decides when to use what accounts to post reviews with what ratings to which targets.
When multiple strategies are used at the same time, the mixed attack strategy is denoted by $A(\mathsf{p})=\mathbb{E}_{k\sim \mathsf{p}} [a_k]=\sum_{k=1}^{K} (p_k a_k)$,
where $\mathsf{p}=[p_1,\dots, p_K]$
is the mixture parameter with $\sum p_k=1$.\footnote{Rigorously, $A(\mathsf{p})$ is not a weighted sum of numbers, since $a_k$ are not numbers but a spam-generating function. Instead, think of $A(\mathsf{p})$ as a weighted sum of functions.}

No single detector can tackle all spamming strategies, and a myriad of detectors based on different spamming assumptions have been proposed in~\cite{Akoglu2013,Rayana2015, Hooi2016,Xie2012a, Mukherjee:2013uk}. See the related work for more details.
Ensembling a diverse set of detectors has been proven useful~\cite{Rayana2015, ren2019ensemfdet}.
We assume there is an array of $L$ base detectors $\mathsf{d}=[d_1,\dots, d_L]$ and each detector outputs a number in $[0,1]$,
with a larger value denoting more suspiciousness.
The detectors are given various importance $\mathsf{q}=[q_1,\dots, q_L]$, so that the detector in effect is $D({\mathsf{q}})=\sum_{l=1}^{L} (q_l d_l)=\mathsf{q}^\top\mathsf{d}$.

\noindent \textbf{Challenges and the threat model}. While the existing detectors are designed to defend against one or multiple spamming strategies,
the underlying \textit{true} spamming strategy has never been fully revealed.
To make the research problem tractable,
we are not trying to discover the \textit{true} spamming strategy, but instead addressing the strongest spamming strategy that the spammers can engineer given due knowledge and resources~\cite{uesato2018adversarial}.
This goal is more practically meaningful:
professional spammers can access the details of spam detectors through published papers~\cite{Mukherjee:2013uk},
reverse engineering from labeled spams~\cite{Rayana2015}, and detection explanations~\cite{reviewmeta}, and update their spamming strategies to evade a fixed detector.
Indeed,
previous work~\cite{Hooi2016,Shah2014,Chen2017,Yao:2017bd} have considered spams in camouflage for evading a \textit{fixed} detector.
Even if the fixed detector remains unknown to the spammer,
there are only a handful of salient detection signals,
leading to a relatively small space of detectors (all $\mathsf{q}$ over the set of known salient detectors).
A spammer can continue to probe the current detector until it finds something vulnerable and then crafts a strong spamming strategy.

The second challenge is that,
spam detection is treated as a binary classification problem and
evaluated using accuracy~\cite{Hooi2016}, recall~\cite{mukherjee2013spotting}, precision, AUC, or nDCG~\cite{Rayana2015, Wang2017}.
These metrics are not well aligned with the practical spamming goal, namely, to perturb product reputation and revenue in a relatively short period.
We observed that the amount of perturbation is only weakly correlated with the detection recall score.
See Figures~\ref{fig:pe_f1} and \ref{fig:yelpync_zip}.
In reality, human spam screening and removals are limited to the top-$k$ suspicious spams,
and the revenue manipulation is not entirely due to the \textit{number} of missed spams (false negatives).
We identify the following vulnerability:
massive obvious spams can easily be detected, saturate the human screening capacity, and pose a deceivingly high detection accuracy,
while a small number of spams posted by elite accounts are missed by the top-$k$ screening and remain to manipulate the target reputation and revenue significantly~\cite{luca2016reviews}.
We term such detectors as ``accuracy-focusing detectors'', as they optimize accuracy and other related metrics (F1-score, recall rate, etc.)

\section{Methodology}
\label{sec:method}
\subsection{Turning Reviews into Business Revenues}
\label{sec:revenue}
A high (or low) rating will contribute to the growth (or decrease) in the sales of a product~\cite{luca2016fake,forman2008examining},
and a one-star increase in average rating contributes to a 5 to 9 percent increase in revenues~\cite{luca2016reviews, lappas2016impact}.
More specifically,
the unit increase in revenue can be attributed to two different types of reviews:
those posted by regular accounts (``regular reviews'') and those by elite accounts (``elite reviews'').
Different from the regular accounts, an elite account is recognized by the review system
if the account has frequently posted high-quality reviews.
The elite reviews are more influential in product revenue~\cite{luca2016reviews}, as they count more in product evaluation in the review system and are more frequently presented to the customers than the regular reviews.

Formally,
let $\mathcal{R}$ ($\mathcal{R}_E$, resp.) be the set of all (elite, resp.) reviews before a spamming attack,
and let
$\mathcal{R}(v)$ ($\mathcal{R}_E(v)$, resp.) be the (elite, resp.) reviews posted towards product $v$.
We adopt the \textit{revenue estimation function} $f(v; \mathcal{R})$ from~\cite{luca2016reviews}
to measure the influence of the reviews on the revenue of product $v$:
\begin{align}
    f(v; \mathcal{R})&= \beta_{0}\times \RI(v;\mathcal{R}) + \beta_{1}\times \ERI(v;\mathcal{R}_E(v)) + \alpha\nonumber\\
    &=\beta_{0}\times(g(\mathcal{R}(v)) - g(\mathcal{R})) + \beta_{1}\times g(\mathcal{R}_E(v)) + \alpha,
\label{eq:revenue}
\end{align}
where the function
$g(\cdot)$ computes the average rating from the given set of reviews.
$\RI(v; \mathcal{R})\triangleq g(\mathcal{R}(v)) - g(\mathcal{R})$
is the \textbf{R}eview \textbf{I}nfluence derived from how much the average rating of $v$ is better or worse than the overall average rating.
$\ERI(v;\mathcal{R}_E(v))\triangleq g(\mathcal{R}_E(v))$
is the \textbf{E}lite \textbf{R}eview \textbf{I}nfluence due to the average rating contributed solely by the elite reviews for $v$.
$\beta_{0}$ and $\beta_{1}$ are coefficients of the two influences,
and
$\alpha$ is the baseline revenue of all items\footnote{
$\beta_{0}=0.035$, $\beta_{1}=0.036$, and $\alpha=1$ based on the empirical findings in~\cite{luca2016reviews}.
$\beta_{1}g(\mathcal{R}_E(v))$ is much larger than $\beta_{0}(g(\mathcal{R}(v)) - g(\mathcal{R}))$ since $\beta_0$ multiplies a difference.}.
Although these coefficients are estimated using Yelp's data and may not be applicable to other review systems, the underpinning of the paper can be applied to other systems such as Amazon so long as this estimation can be done.

\subsection{Practical Spamming and Detection Goals}
\label{sec:practical_metric}
While the spammers aim to manipulate (promote or demote) the targets' revenues,
the spam detector aims to tame such manipulations.
In the following, assuming a spammer wants to promote the given targets' revenue, we define the goals of the spammer and the defender (demotions are discussed at the end of Section~\ref{sec:method}).
We denote a spamming strategy by $A(\mathsf{p})$ with parameters $\mathsf{p}$,
and the detector with detection strategy $\mathsf{q}$ by  $D(\mathsf{q})$.
The letter $v$ is designated to a target product.

\vspace{.05in}
\noindent \textbf{A practical spamming goal.}
Let $\mathcal{R}(\mathsf{p})$ be the spams posted using the mixed spamming strategy $A(\mathsf{p})$,
and $\mathcal{R}(\mathsf{p}, \mathsf{q})$ be the false-negative spams that remain after $\mathcal{R}(\mathsf{p})$ is purged by the detector $D(\mathsf{q})$~\footnote{Possibly with human screening of the detected reviews to further reduce false positives}.
Based on the revenue $f$ defined in Eq. (\ref{eq:revenue}),
a metric to evaluate the Practical Effect (PE)
of spamming using $A(\mathsf{p})$ on $v$
against the detection of $D(\mathsf{q})$ is:
\begin{equation}
\begin{split}
      \PE(v;\mathcal{R}, \mathsf{p}, \mathsf{q})
    &= f(v;\mathcal{R}(\mathsf{p}, \mathsf{q})) - f(v;\mathcal{R})\\
    &= \beta_{0}\times\Delta\RI(v) + \beta_{1}\times\Delta\ERI(v),
\end{split}
\label{eq:practical_metric}
\end{equation}
which is the difference in the revenues of $v$
after and before the spamming and the detection.
\begin{align}
\Delta\RI(v) & = \RI(v;{\mathcal{R}(\mathsf{p}, \mathsf{q})}) - \RI(v; \mathcal{R})=g(\mathcal{R}(v;\mathsf{p}, \mathsf{q}))-g(\mathcal{R}(v))\nonumber\\
\Delta\ERI(v) & = \ERI(v;\mathcal{R}_E(\mathsf{p}, \mathsf{q})) - \ERI(v;\mathcal{R}_E)=g(\mathcal{R}_E(v;\mathsf{p}, \mathsf{q}))-g(\mathcal{R}_E(v))\nonumber
\end{align}
are the change in influences due to the missed spams.

$\PE(v)$ can be negative when the spams and the detector bring $v$'s rating down.
This can happen when the human inspectors delete some organic reviews with high ratings from $v$.
We assume that the inspectors will screen the top $k$ ($k$ is typically small) detected reviews carefully so that there is no genuine review deleted.
A drop in revenue can also be caused by a spamming strategy that posts negative reviews for the camouflage purpose.
We consider such more complex strategies in future work.
\begin{tcolorbox}[
enhanced,clip upper,
colframe=gray,colback=white,boxrule=1pt,arc=3pt,
boxsep=2pt,left=-2pt,right=3pt,top=3pt,bottom=3pt
]
\begin{equation}
\textbf{Spamming goal:}\hspace{.4in}
    \max_{\mathsf{p}}\hspace{.1in}\max\{0, \PE(v; \mathcal{R}, \mathsf{p}, \mathsf{q}))\}.
\end{equation}
\end{tcolorbox}

\vspace{.05in}
\noindent \textbf{A practical detection goal.}
The practical detection goal should be
minimizing $\max\{0, \PE(v; \mathcal{R}, \mathsf{p}, \mathsf{q}))\}$ and make sure the resulting detection strategy $q$ will not entirely focus on detection accuracy
but will suppress the spamming promotion.

In the following, we define a back-propagation algorithm for detection strategy learning.
According to Eq.~\ref{eq:practical_metric},
$\max\{0, \PE(v; \mathcal{R}, \mathsf{p}, \mathsf{q}))\}$ summarized the effect of the false-negative spams $\mathcal{R}(\mathsf{p}, \mathsf{q})$.
To guide the detection strategy $q$,
we first back-propagate (or attribute) $\max\{0, \PE(v; \mathcal{R}, \mathsf{p}, \mathsf{q}))\}$ to individual false-negative spams, and
the attributed costs are further back-propagated to the current detection strategy $\mathsf{q}$ that lead to the false negatives.
From the spammer's perspective,
elite spamming reviews are more influential.
From the detector's perspective,
a missed elite spamming review leads to a larger amount of
revenue manipulation that a missed regular spamming review.
Based on cost-sensitive learning~\cite{elkan2001foundations},
we turn the spamming effect $\max\{0, \PE(v; \mathcal{R}, \mathsf{p}, \mathsf{q}))\}$ into detection costs according to different detection outcomes:
the detection costs of true positives and true negatives are 0 ($C_{\textnormal{TP}}=C_{\textnormal{TN}} = 0$);
the false positives will be handled by human screening and will cause zero effect on the product revenue ($C_{\textnormal{FP}}=0$);
a false negative will not contribute to the promotion of product $v$ if PE$(v)\leq 0$ unless otherwise it will contribute to the following cost through $\Delta\RI(v)$ and $\Delta\ERI(v)$:
\begin{equation}
    C_{\textnormal{FN}}(v,r) = 
    \frac{\beta_{0}\Delta\RI(v)}{Z_1}
    +\mathds{1}_{r\in \mathcal{R}_E(\mathsf{p}, \mathsf{q})}\left[\frac{\beta_{1}\Delta\ERI(v)}{Z_2}\right],
\label{eq:detector_cost}
\end{equation}
where $Z_1$ and $Z_2$ are respectively the amount of non-elite and elite spams posted towards $v$.
$\mathds{1}$ is the indicator function.
Based on the analysis, the detection goal is defined as:

\begin{tcolorbox}[
enhanced,clip upper,
colframe=gray,colback=white,boxrule=1pt,arc=3pt,
boxsep=2pt,left=-1pt,right=3pt,top=3pt,bottom=3pt
]
\begin{equation}
\textbf{Detection goal:}\hfill
    \min_{\mathsf{q}}\hspace{.1in}\mathcal{L}(\mathsf{q})
     =  \frac{1}{|\mathcal{R}(\mathsf{p}, \mathsf{q})|}\sum_{r \textnormal{ is FN}}C_{\textnormal{FN}}(v,r),
\end{equation}
\end{tcolorbox}
\noindent where $v$ is the target product that $r$ was posted to.
$C_{\textnormal{FN}}(v,r)$ implicitly depends on the strategy $\mathsf{q}$, through the ranking of reviews by the detector $D(\mathsf{q})$ and the screening of the top $k$ reviews.
To facilitate the optimization of $\mathsf{q}$,
we define
the following \textit{cost-sensitive} surrogate detection loss function:
\begin{equation}
    \mathcal{L}_{\mathsf{q}} 
     =  \frac{1}{|\mathcal{R}(\mathsf{p}, \mathsf{q})|}\sum_{r \textnormal{ is FN}} - C_{\textnormal{FN}}(v,r) \log P(y=1 | r;\mathsf{q}),
\label{eq:cost_sensitive_loss}
\end{equation}
where $y\in\{0,1\}$ is the label of $r$ ($y=1$ if and only if $r$ is spam).
$P(y=1|r;\mathsf{q})$ is the probability of $r$
being a spam predicted by $D(\mathsf{q})$:
\begin{equation}
    \label{eq:ensemble}
    P(y=1 | r;\mathsf{q})=\sigma(\mathsf{q}^\top \mathsf{d}(r)),\hspace{.1in} \mathsf{d}(r)=[d_1(r),\dots, d_L(r)],
\end{equation}
where $\sigma$ represents the sigmoid function, the surrogate loss says to reduce the cost $C_{\textnormal{FN}}(r,v)$, the detector should output a large $P(y=1 | r;\mathsf{q})$ so that the spam $r$ can be pushed into the top $k$ suspicious reviews.

\subsection{Minimax Game and Optimization}
\label{sec:game}

The spammer and the detector's goals are now well-aligned:
the spammer aims at promoting the revenues of the targets $v\in\mathcal{V}_T$ and the detector wants to suppress such promotion.
They will play the following zero-sum game over the practical spamming effect $\max\{0, \PE(v; \mathcal{R}, \mathsf{p}, \mathsf{q}))\}$:
\begin{equation}
\label{eq:game}
    \min_{\mathsf{q}}\max_{\mathsf{p}} \hspace{.1in} \sum_{v\in \mathcal{V}_T}\max\{0, \textnormal{PE}(v; \mathcal{R}, \mathsf{p}, \mathsf{q})\}.
\end{equation}
Solving the above game will lead to a detector that can withstand any mixing spamming strategies weights $[a_1,\dots, a_K]$.
In particular,
we aim at a robust detector, parameterized by $\mathsf{q}^\ast$, that will minimize practical spamming effects caused by any spamming strategies $A(\mathsf{p})$.
One limitation of $\mathsf{q}^\ast$ is that,
during test time,
a spammer may use a pure strategy not considered in the pure strategies $[a_1,\dots, a_K]$ that $D(\mathsf{q}^\ast)$ was trained on.
It is thus essential to include representative spamming strategies during the training of $\mathsf{q}$.
Exhausting all spamming strategies is out of the scope of this work, and in Section~\ref{sec:attacks_detectors}, we specify evasive attacks against mainstream detectors.
In the sequel, ``spammer'' can refer to the fictitious spammers during training or the real spammer during test time.

The objective function is not differentiable, as the calculation of PE (shown in Eq. (\ref{eq:practical_metric})) are based on nonlinear operators such as adding reviews to $\mathcal{R}$ using the spamming strategy $A(\mathsf{p})$, ranking the reviews in $\mathcal{R}(\mathsf{p})$ based on $P(y=1|r;\mathsf{q})$ (which depends on the detectors $[d_1,\dots, d_L]$), the removal of spams from the top $k$ suspicious ones, and the consideration of elite accounts.
We note that such non-differentiability is a necessary evil in robust spam detection: diverse spamming and detection strategies help explore the limit of both players, while many state-of-the-art spam attacks and detectors are non-differentiable, non-continuous~\cite{Yao:2017bd}, stochastic~\cite{Hooi2016}.
Therefore, gradient-based optimization methods are not applicable, and we propose a multi-agent non-cooperative reinforcement learning approach~\cite{Tan93,Hu2003,Littman1994} and use stochastic optimization based on Monte-Carlo to address the challenges.
Regarding the spammer and the detector as two competing agents who play the attack and defense game in Eq. (\ref{eq:game}).
The experiences of detector and spammer will be obtained from a $T$-step roll-out of
two \textit{dependent} Markov Decision Processes (MDPs)~\cite{Sutton1998book}:
for each episode (indexed by $t=1,\dots, H$) of the $H$ episodes of game playing,
the spammer will attack the review system using the current mixed strategy $A(\mathsf{p}^{(t)})=\sum_{k=1}^{K} (p_k^{(t)} a_k)$ and the detector will respond with the mixed detection strategy $D(\mathsf{q}^{(t)})=\sum_{l=1}^{L} (q_l^{(t)} d_l)$.
Both mixing parameters $\mathsf{p}$ and $\mathsf{q}$ will be updated as follows.

\vspace{.1in}
\noindent\textbf{An MDP for the spammer}.
We adopt the multi-armed bandit formulation~\cite{BubeckC12} so that the spammer has no \textit{explicit} state representation of the review system but acts based on the practical spamming effects.
To be precise,
the spammer maintains the distribution $\mathsf{p}$ over the \textit{base} spamming strategies $a_1,\dots, a_K$ as a policy.
In episode $t$, for each target $v$,
the spammer samples an action $a_k$ with $k\sim \texttt{Multinominal}(\mathsf{p}^{(t)})$ to execute a spamming attack (via posting a fake review to $v$ using an account selected by $a_k$).
The attack on $v$ by $a_k$ is denoted by $a_k\rightsquigarrow v$.
At the end of the episode, the detector uses the mixed strategy $D(\mathsf{q}^{(t)})$ to remove some posted spams, and the final PE$(v;\mathcal{R}, \mathsf{p}^{(t)}, \mathsf{q}^{(t)})$ is calculated as in Eq. (\ref{eq:practical_metric}).
The reward for the spamming strategy $a_k$ is the portion of PE in this episode due to the spams posted by $a_k$ but missed by $D(\mathsf{q}^{(t)})$.
The specific credit assignment at episode $t$ is:
\begin{equation}
    G^{(t)}(a_k) =\sum_{a_k\rightsquigarrow v}
    \sigma\left(
    \frac{\textnormal{PE}(v;\mathcal{R}, \mathsf{p}^{(t)}, \mathsf{q}^{(t)})- \textnormal{AVG(PE)}}{Z}\right),
\label{eq:attack_rewards}
\end{equation}
where $Z=\max_{v\in\mathcal{V}_T}\textnormal{PE}(v;\mathcal{R}, \mathsf{p}^{(t)}, \mathsf{q}^{(t)})-\min_{v\in\mathcal{V}_T}\textnormal{PE}(v;\mathcal{R}, \mathsf{p}^{(t)}, \mathsf{q}^{(t)})$.
The maximum, minimum, and average are calculated over all targets $\mathcal{V}_T$,
including those not attacked by the strategy $a_k$.
The subtraction of the average from PE$(v;\mathcal{R}, \mathsf{p}^{(t)}, \mathsf{q}^{(t)})$ can help differentiate effective attacks from less effective ones~\cite{Sutton1998book}.
The rewards are accumulated across multiple episodes of the game, and $\mathsf{p}^{(t)}$ is as:
\begin{equation}
    \label{eq:p_update}
    p_k^{(t+1)}\propto \left[p_k^{(0)} + \eta  \sum_{\tau=1}^{t}
    \frac{G^{(\tau)}(a_k)}{|\{v:a_k\rightsquigarrow v\}|} \right].
\end{equation}


\vspace{.1in}
\noindent\textbf{An MDP for the detector}.
In episode $t$,
the detector senses the state of the review system as the vector 
$\mathsf{d}(r_i)=[d_1(r_i),\dots, d_L(r_i)]$ for each review $r_i\in \mathcal{R}(\mathsf{p}^{(t)})$ after the spamming attack $A(\mathsf{p}^{(t)})$.
To evaluate the current detection strategy, the labels of the added spams are not disclosed to the detector when the detector $D(\mathsf{q}^{(t)})$ takes its actions.
To simplify training, the base detectors $d_l$ are fixed.
The strategy $\mathsf{q}^{(t)}$ generates the probability $P(y=1 | r;\mathsf{q})$ on each review $r\in \mathcal{R}(\mathsf{p}^{(t)})$ according to Eq.~(\ref{eq:ensemble}).
The top-$k$ suspicious reviews based on $P(y=1|r,\mathsf{q}^{(t)})$ are removed, leading to the set $\mathcal{R}(\mathsf{p}^{(t)},\mathsf{q}^{(t)})$,
which potentially contains false negatives that contribute to PE$(v;\mathcal{R}, \mathsf{p}^{(t)}, \mathsf{q}^{(t)})$.
The spamming effect attributed to individual false negatives
is $C_{\textnormal{FN}}$ defined in Eq. (\ref{eq:detector_cost}).
The mixing parameter $\mathsf{q}^{(t)}$ will be updated to $\mathsf{q}^{(t+1)}$ by minimizing the following cost-sensitive loss function:
\begin{equation}
    \mathcal{L}(\mathsf{q}) = \frac{1}{|\mathcal{R}(\mathsf{p}^{(t)},\mathsf{q}^{(t)})|} \sum_{r\in\mathcal{R}(\mathsf{p}^{(t)},\mathsf{q}^{(t)})}-C_{\textnormal{FN}}(v ,r)\log P(y=1|r;\mathsf{q}),
\label{eq:detector_loss}
\end{equation}
where $v$ is the target that $r$ was posted to.
After the detection, the current episode is done, and both agents move to the next episode and play the same game with the updated parameters $(\mathsf{p}^{(t+1)}, \mathsf{q}^{(t+1)})$.



\vspace{.1in}
\noindent \textbf{Optimization algorithm}.
We propose Nash-Detect in Algorithm~\ref{alg:defense} for finding a Nash equilibrium $(\mathsf{p}^\ast, \mathsf{q}^\ast)$,
and $\mathsf{q}^\ast$ is the resulting robust detection strategy. Figure~\ref{fig:motivations} presents a toy-example of Nash-Detect.
At a Nash equilibrium, both the spammer and detector will not want to change their strategies $(\mathsf{p}^\ast, \mathsf{q}^\ast)$
since it will not lead to further benefit.
There can be multiple Nash equilibria and the algorithm just finds one.
It is challenging to prove the uniqueness of Nash equilibria except in some very restrictive problem settings.
Experiments show that Nash-Detect always finds a robust detector regardless of what $\mathsf{p}^\ast$ is. 
Nash-Detect is trained in $H$ episodes.
During each episode $t=1,\dots, H$, there are the forward and backward steps:
\begin{itemize}[leftmargin=*,labelindent=1mm,labelsep=1mm]
    \item 
\underline{Inference (forward)}: for each target $v\in\mathcal{V}_T$, the fictitious spammer samples one pure strategy from $a_1,\dots, a_K$ according to $\mathsf{p}^{(t)}$ and
posts spams to $v$ using an account determined by the selected strategy.
The sampling of $a_k$ and the selection of accounts by $a_k$ are conducted independently among the targets.
That is, posting spams to one target will not affect the selection of $a_k$ and how $a_k$ runs its attack on the subsequent targets.
It is left as future work to
consider updating the review data before spamming the next target so there can be dependencies between the spams towards two targets. 
When all targets receive their spams,
the detector $D(\mathsf{q}^\ast)$ removes the top-$k$ suspicious spams and the practical effect $\PE(v;\mathcal{R}, \mathsf{p}, \mathsf{q})$ is computed as the final output of the inference (forward pass).
\item \underline{Learning (backward)}:
the current rewards for the spamming strategies are calculated using Eq. (\ref{eq:attack_rewards}) and $\mathsf{p}^{(t)}$ is updated to $\mathsf{p}^{(t+1)}$.
Simultaneously,
the detection strategy $\mathsf{q}^{(t)}$ is also updated by minimizing the loss function Eq. (\ref{eq:cost_sensitive_loss}) to obtain $\mathsf{q}^{(t+1)}$.
\end{itemize}

\begin{algorithm}
\caption{Nash-Detect: Training a Robust Spam Detector}
\label{alg:defense}
\begin{algorithmic}[1]

\State \textbf{Input}: all reviews $\mathcal{R}$, target items $\mathcal{V}_{T}$, pure attack strategies $[a_1,\dots, a_K]$, pure spam detectors $[d_1,\dots, d_L]$, initial spamming strategy $\mathsf{p}^{(0)}=[p_1,\dots, p_K]$ and initial detection strategy $\mathsf{q}^{(0)}=[q_1,\dots, q_L]$ to uniform distributions.

\State \textbf{Output}: a Nash equilibrium $(\mathsf{p}^{*}, \mathsf{q}^{*})$.

\Repeat\Comment{Go through the $H$ episodes indexed by $t$}
\State  \underline{Inference}:
\State $\mathcal{R}(\mathsf{p}^{(t)})=\mathcal{R}$.

\ForAll{$v\in\mathcal{V}_{T}$}\Comment{Post fake reviews}

\State Sample $a_k$ using $\epsilon$-greedy for $v$ according to $\mathsf{p}^{(t)}$.

\State 
Post spams to $v$ using $a_k$. 



\EndFor


\State 
Remove spams in the top $k$ reviews detected by $D(\mathsf{q}^{(t)})$.

\State Compute $\PE(v, \mathcal{R}, \mathsf{p}^{(t)}, \mathsf{q}^{(t)})$ using Eq. (\ref{eq:practical_metric}) on $\mathcal{R}(\mathsf{p}^{(t)}, \mathsf{q}^{(t)})$.





\State \underline{Learning}:

\State Compute $C_{\textnormal{FN}}(v,r)$ using Eq. (\ref{eq:detector_cost}) and $G(a_k)$ using~Eq. (\ref{eq:attack_rewards}).


\State Update $\mathsf{p}^{(t)}$ to $\mathsf{p}^{(t+1)}$ using the gains $G(a_k)$.

\State Update $\mathsf{q}^{(t)}$ to $\mathsf{q}^{(t+1)}$ by minimizing Eq. (\ref{eq:detector_loss}).

\Until{$\mathcal{L}(\mathsf{q})$ converges}
\end{algorithmic}
\end{algorithm}

\vspace{.05in}
\noindent \textbf{Discussion}.
Nash-Detect assumes the spammer aims at promoting, rather than demoting the targets.
Nash-Detect can handle demotion by changing the practical spamming effect from $\max\{0, \PE(v; \mathcal{R}, \mathsf{p}, \mathsf{q}))\}$ to $\min\{0, \PE(v; \mathcal{R}, \mathsf{p}, \mathsf{q}))\}$ where $\PE(v; \mathcal{R}, \mathsf{p}, \mathsf{q}))$ shall be made a small negative number to bring down the targets' revenues.

When applying the resulting detector, admittedly, one must know whether a spammer will promote or demote a product.
A simple solution is to learn about the probability that a product will commit self-promotion and apply Nash-Detect trained based on $\max\{0, \PE(v; \mathcal{R}, \mathsf{p}, \mathsf{q}))\}$. For example, pieces of evidence showed that when a product recently received negative reviews,
it is more likely to commit spamming for self-promotion.
On the other hand,
demotions are more likely among competitors who offer similar products (e.g., Samsung vs. Apple~\cite{samsung_apple}, or Starbucks vs. Dunkin' Donuts~\cite{luca2016fake}).
We leave as future work regarding apply two robust detectors to handle demotions and promotions simultaneously.








\subsection{Base Detection and Spamming Strategies}
\label{sec:attacks_detectors}
\noindent \textbf{Base detection strategies.} As mentioned in Section~\ref{sec:prelim}, there have been a variety of graph-based and behavior-based detectors. We select the following five base detectors:
\begin{itemize}[leftmargin=*]
\setlength\itemsep{0.05em}
    \item GANG~\cite{Wang2017}: a social network Sybil detection algorithm via linearized belief propagation.

    \item SpEagle~\cite{Rayana2015}: an advanced belief propagation algorithm verified on Yelp review spam datasets. 

    \item fBox~\cite{Shah2014}: an SVD-based algorithm that spots small-scale suspicious links in social networks.

    \item Fraudar~\cite{Hooi2016}: a fraudster detection algorithm that detects dense blocks in graphs.

    \item Prior: an algorithm that ranks spams based on multiple suspicious behavior features listed in~\cite{mukherjee2013spotting, Mukherjee:2013uk, Rayana2015}.
\end{itemize}
These detectors are unsupervised~\cite{Rayana2015,Hooi2016,Shah2014}, have no or very few parameters~\cite{Hooi2016,mukherjee2013spotting}, are scalable~\cite{Wang2017}, diverse, and representative of explicit assumptions about the spamming strategies, making the resulting detector more robust and interpretable. 
Supplement~\ref{sec:detector_details} shows their implementation details.
There is no hurdle for Nash-Detect to include deep learning methods~\cite{liu2020alleviating,kaghazgaran2019wide} as base detectors.
However, the underlying spamming strategies are \textit{learned} from many \textit{labeled} data, leading to a less interpretable detector (in the sense of explaining the interplay between any two detection and spamming strategies, and why the resulting detector is robust). 
 

\vspace{.05in}

\noindent \textbf{Base spamming strategies.}
Different from the secure machine learning literature~\cite{biggio2013security},
a spamming strategy cannot manipulate some important detection features directly but has to do so via posting new reviews.
On some review systems, a spammer can retract posted reviews and can be used in attacks against clustering and classification on graphs~\cite{zugner2018adversarial,sun2018adversarial}.
We assume that only additions of reviews are allowed,
similar to the attacks on graphs proposed in~\cite{Chen2017}.
The reasons are that adding reviews is the most direct way to perturb target ratings, while deleting reviews can make a spamming account look suspicious~\cite{reviewmeta} and reduce the effects of previous attacks.
There are attacks based on text generation~\cite{Yao:2017bd, kaghazgaran2018combating} and control of spamming tempo~\cite{Ge2018}, but we do not consider such controls in this work for simplicity.
All these prior attacks do not distinguish elite and regular accounts and are not trained to maximize the practical spamming effect.
We consider the situation where elite and regular accounts contribute differently to the practical spamming effect~\cite{zheng2017smoke, Xie2012a}.
We propose the following \textit{base} spamming strategies
that differ in their target detectors and whether elite accounts are employed,
and let Nash-Detect learn the importance of each base spamming strategy. The implementation details of following strategies are shown in Supplement~\ref{attack_details}.
\begin{itemize}[leftmargin=*]
\setlength\itemsep{0.05em}

\item \texttt{IncBP}: it uses elite accounts and tries to evade detectors that use both behavior~\cite{mukherjee2013spotting, Mukherjee:2013uk, Xie2012a} and
graph information~\cite{Rayana2015,Wang2017}.
It employs linearized belief propagation (LinBP)~\cite{Wang2017} on a Markov Random Field (MRF) to estimate account suspiciousness.
The strategy estimates the suspiciousness of controlled accounts after posting fake reviews to one target and posts reviews to the next target using accounts with the minimum suspiciousness.

\item \texttt{IncDS}: it uses elite accounts and aims to evade detectors that regard dense blocks suspicious~\cite{Hooi2016}.
\texttt{IncDS} estimates the suspiciousness of each account using the density of the subgraph composed of its neighbors after a spam is posted. \texttt{IncDS} selects the account with minimum density to post the next spam.

\item \texttt{IncPR}: it uses elite accounts and aims to evade the detector Prior.
The spamming process is iterative with account suspiciousness estimation, similar to \texttt{IncBP} and \texttt{IncDS}.

\item \texttt{Random}~\cite{Shah2014, Hooi2016}: it randomly picks an elite account to post a spam. 

\item \texttt{Singleton}: it uses new accounts, each of which posts only one review.
Such spams can avoid creating dense blocks and thus can evade Fraudar.
\end{itemize}

\section{Experiments}
\label{sec:experiments}


\subsection{Experimental Setup}
\label{sec:exp_settings}

\noindent \textbf{Datasets.}
We conduct experiments on three Yelp review datasets (YelpChi, YelpNYC, and YelpZip) used in~\cite{Mukherjee:2013uk, Rayana2015}.
The datasets contain reviews labeled as spams and legitimate, but we simply assume the set of reviews are $\mathcal{R}$ where the spammer will start their spamming attacks.
Dataset details are in the Supplement.
\vspace{.05in}

\noindent \textbf{Elite account selection.}
The datasets only contain partial information of each account and we could not crawl the accounts' elite status from Yelp.
Since Yelp takes the number of reviews posted by the accounts as a crucial factor in determining account elite status,
and there are estimated $6\%-9\%$ elite members among all accounts of Yelp~\cite{yelpelite, dai2012optimal, kim2015discovering},
we regard accounts with more than \textit{ten} reviews as elite accounts in the experiments,
representing $1.4\%$, $4.30\%$ and $4\%$ of the total accounts on the three datasets, respectively.

\vspace{.05in}

\noindent \textbf{Spamming attack and detector setting.}
We select fixed sets of elite accounts and target products to train/test all spamming and defense strategies (details are in Supplement~\ref{sec:attack_setting}).
After calculating the review suspicious scores using Eq. (\ref{eq:ensemble}),
all reviews are ranked by their suspicious scores in descending order, and we remove the top $k\%$ suspicious reviews as spams. We set $k=1$ in all experiments.

\vspace{.05in}
\noindent \textbf{Evaluation criteria.}
According to Section~\ref{sec:game}, we expect that Nash-Detect can find the optimal detector configuration ($\mathsf{q}^{\ast}$) by playing the minimax game between the spammer and defender.
With $D(\mathsf{q}^{\ast})$, the performance of $A(\mathsf{p})$ should be worse than the worst-case of each single attack against a single detector~\cite{uesato2018adversarial}.
It will show that the detectors configured by Nash-Detect could defend the attacks better than any single detector.
The performance of attacks and defenses are both measured by the practical effect (Eq. (\ref{eq:practical_metric})).
 
\begin{figure}[h]
    \centering
    \begin{subfigure}[b]{0.23\textwidth}
        \centering
        \includegraphics[width=\textwidth]{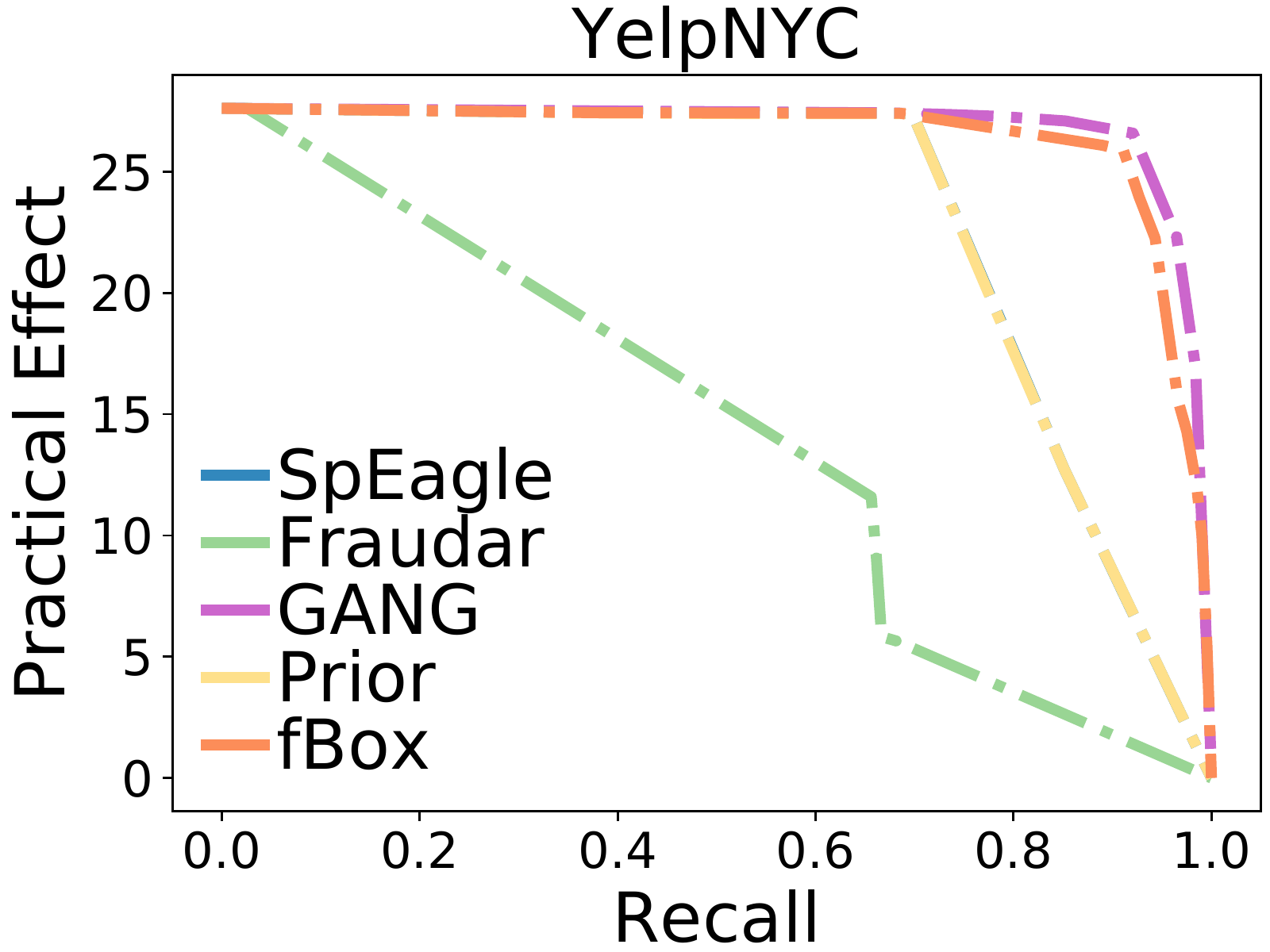}
    \end{subfigure}
    \hfill
    \begin{subfigure}[b]{0.23\textwidth}
        \centering
        \includegraphics[width=\textwidth]{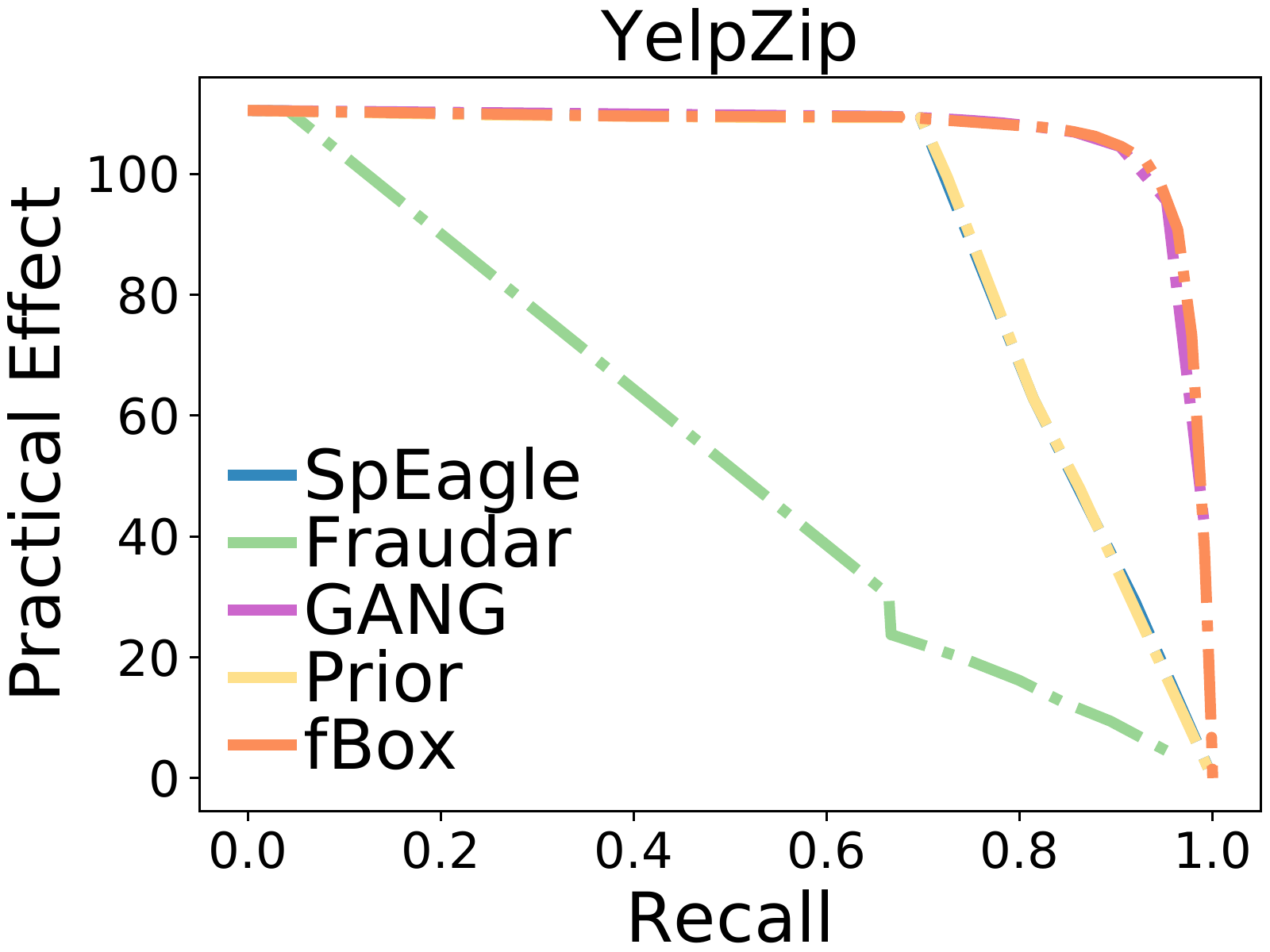}
    \end{subfigure}
    \caption{\footnotesize Practical Effect vs. Recall for different detectors against ensemble attacks on YelpNYC and YelpZip.}
    \label{fig:yelpync_zip}
\end{figure}

\vspace{-.2in}

\subsection{Practical Effect vs. Accuracy}
\label{sec:pe_f1score}
To verify the advantage of the proposed practical goal,
we plot the detection results of five single detectors against five attacks under three datasets.
Figure~\ref{fig:pe_f1} and Figure~\ref{fig:yelpync_zip} show the Practical Effect-Recall curves of five detectors against multiple attacks on three Yelp datasets.
The practical effect is calculated according to Eq. (\ref{eq:practical_metric}) and recall score is obtained by using the suspicious scores as a detector.
Given the review suspicious scores, as we decrease the classification threshold from 100\% to 0\%,
more spams will be detected,
the recall score increase from 0 to 1, and the practical effect should gradually decrease.
However, most detectors (GANG, fBox, SpEagle, Prior) only reduce a small practical effect of attacks when their recall scores increase from 0 to 0.8.
It demonstrates that the practical goal proposed by us could capture the practical effect of the spamming campaign,
and a high accuracy metric does not mean a good practical detection performance. 
We also employ the practical effect to show the vulnerability of individual detectors.
Table~\ref{tab:worst_case} shows the practical effect of individual detectors against individual attacks.
One can see that if the spammer knows that a specific detector (such as Fraudar and fBox) is adopted by the defender,
the spammer can adopt the spamming strategy (such as \texttt{Random} or \texttt{IncBP}) that leads to the most practical effect with respect to the known detector.
Therefore, a detector ensemble configuration $D(\mathsf{q})$ is necessary.

\begin{table}[h]
\small
\centering
\caption{\footnotesize The practical effect of detectors against attacks under YelpChi.}
\resizebox{1\linewidth}{!}{%
\begin{tabular}{|l|c|c|c|c|c|}  
\hline
& GANG &  SpEagle & fBox & Fraudar  & Prior    \\ 
\hline
\texttt{IncBP} & 4.8916 & 4.9052  & 4.9125  & 1.4203& 4.9099  \\
\texttt{IncDS} & 4.9010 & 4.9052  & 4.9110  & 4.8959 & 4.9099\\
\texttt{IncPR} & 4.9010 & 4.9052  & 4.9105  & 3.0716&  4.9099 \\
\texttt{Random} & 4.9010 & 4.9052  & 4.9092  & 4.8962& 4.9099 \\
\texttt{Singleton} & 0.5300 & 0.5865 & 0.5783 & 0.5771 &0.5912\\
\hline
\end{tabular}}
\label{tab:worst_case}
\end{table}

\subsection{Nash-Detect Training Process}
\label{sec:game_outcome}

\begin{figure*}
    \centering
    \begin{subfigure}[b]{0.48\textwidth}
        \centering
        \includegraphics[width=\textwidth]{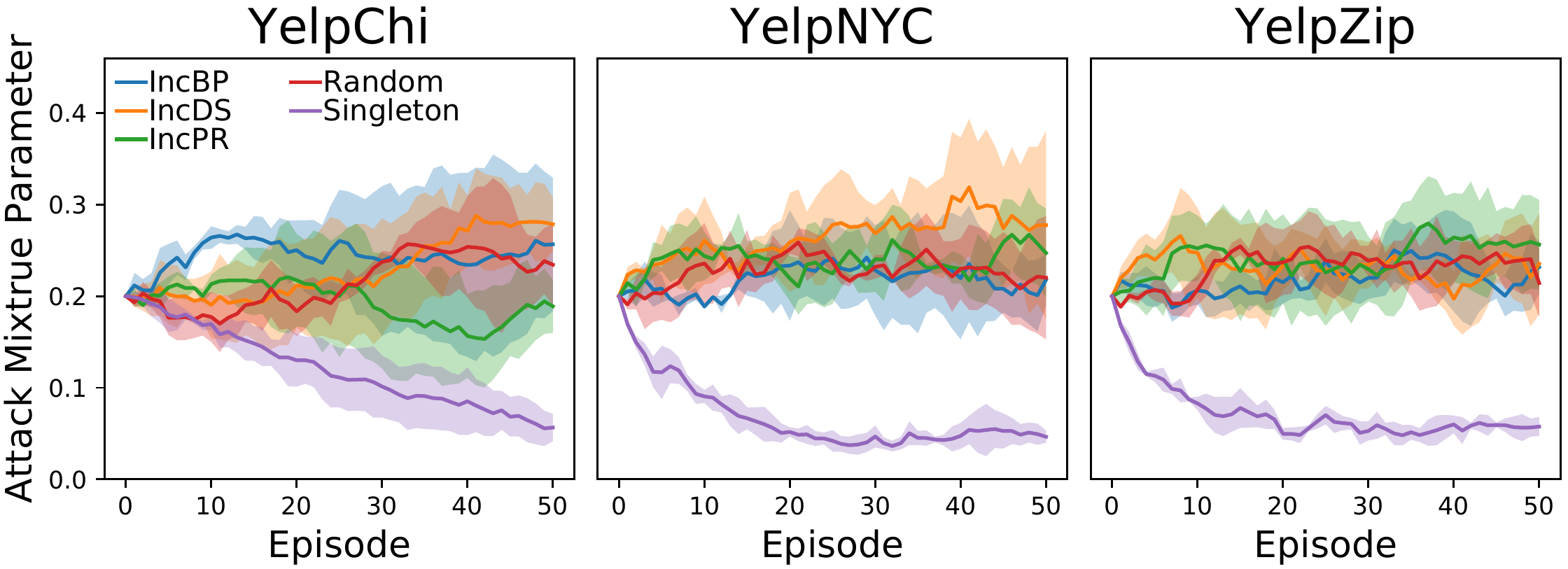}
        \caption{\footnotesize Spamming strategy mixture parameter ($\mathsf{p}$).}
        \label{fig:attack_update}
    \end{subfigure}
    \hfill
    \begin{subfigure}[b]{0.48\textwidth}
        \centering
        \includegraphics[width=\textwidth]{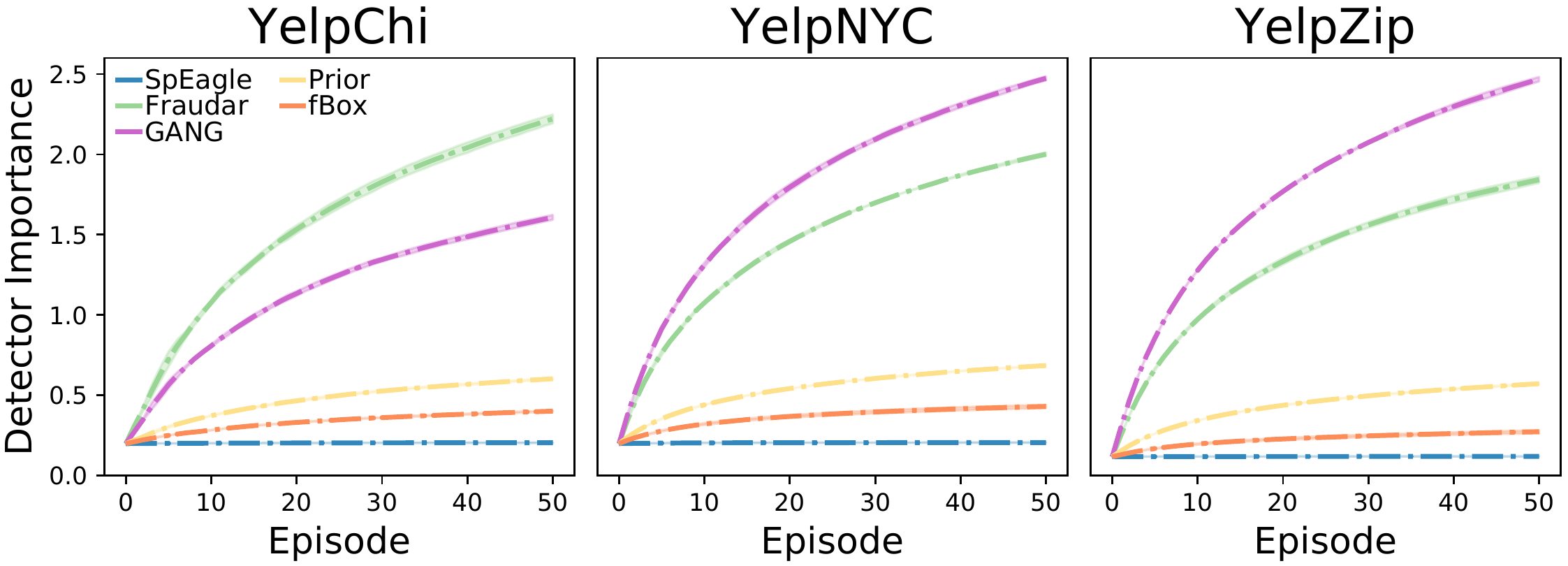}
        \caption{\footnotesize Detector importance ($\mathsf{q}$).}
        \label{fig:detector_update}
    \end{subfigure}
    \bigskip
    \begin{subfigure}[b]{0.48\textwidth}
    \centering
    \includegraphics[width=\textwidth]{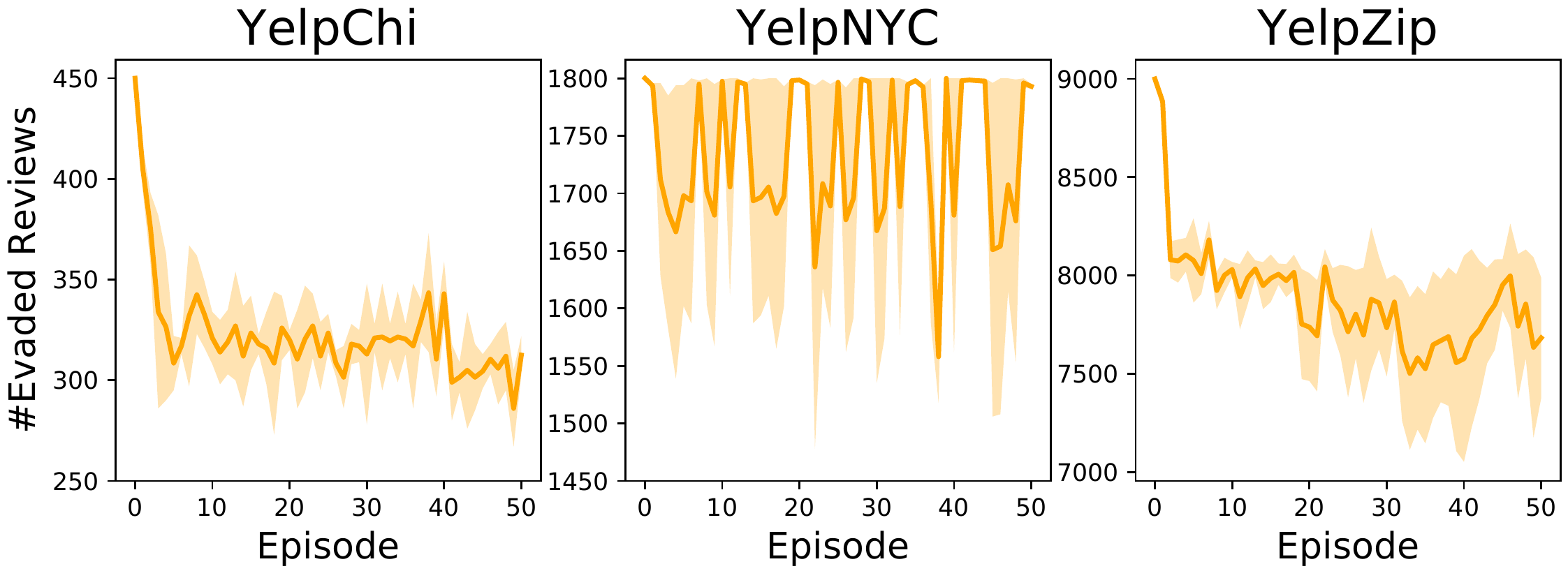}
    \caption{\footnotesize The number of evaded (false-negative) reviews ($\mathcal{R}(\mathsf{p}, \mathsf{q})$).}
    \label{fig:evaded_reviews}
    \end{subfigure}
    \hfill
    \begin{subfigure}[b]{0.48\textwidth}
    \centering
    \includegraphics[width=\textwidth]{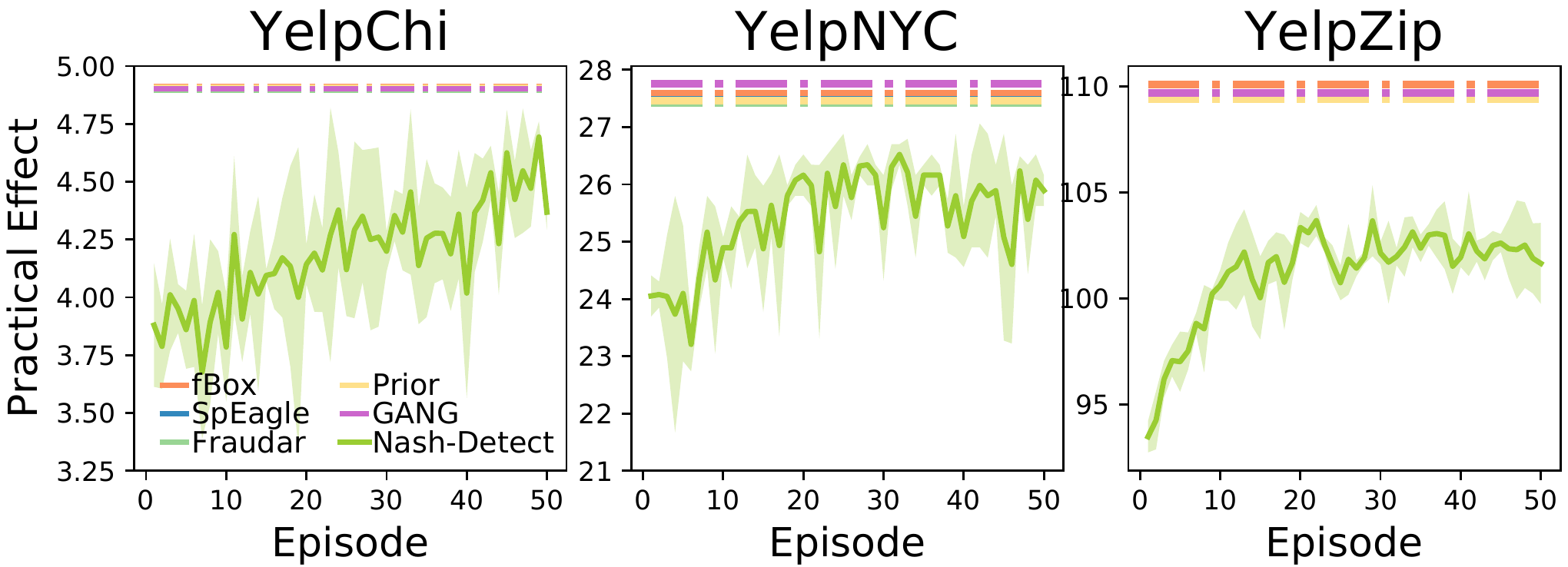}
    \caption{\footnotesize PE of Nash-detect (green) vs. worst-case PE of the single detector.}
    \label{fig:practical_effect}
    \end{subfigure}
    \bigskip
    \begin{subfigure}[b]{0.48\textwidth}
    \centering
    \includegraphics[width=\textwidth]{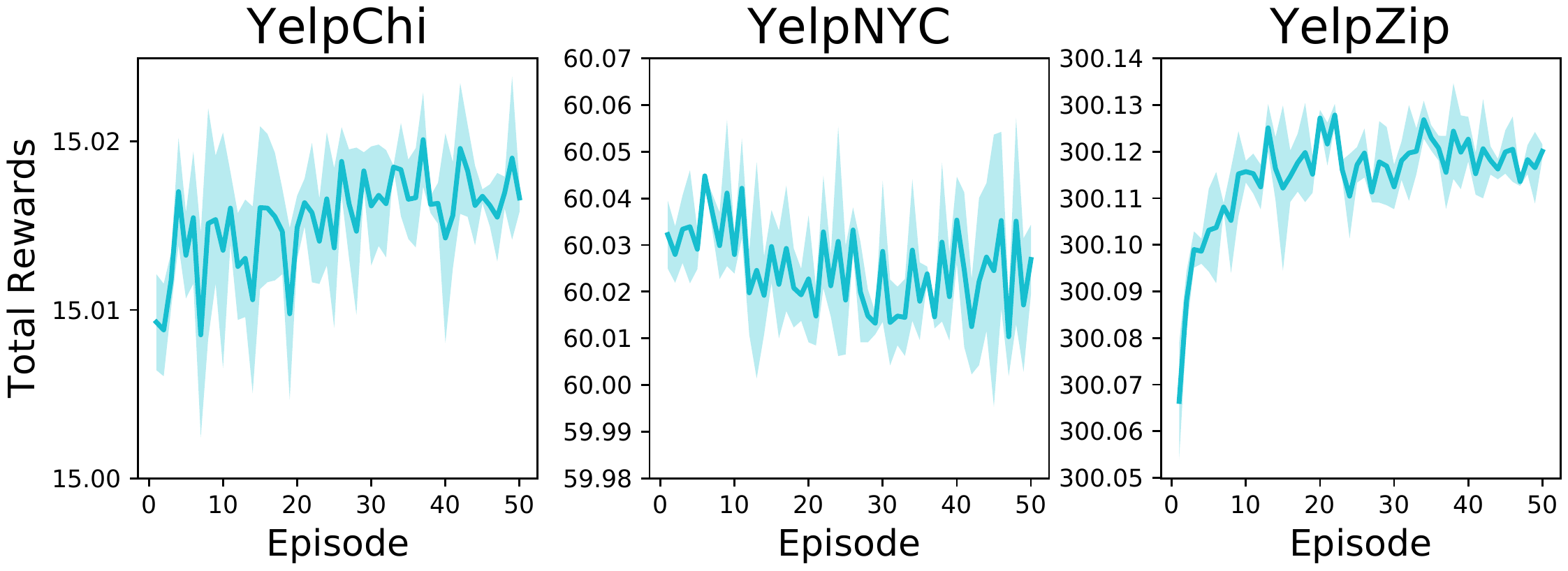}
    \caption{\footnotesize The total rewards for all spamming strategies ($\sum_{a_{k}\in A}G(a_{k})$)}
    \label{fig:total_return}
    \end{subfigure}
    \hfill
    \begin{subfigure}[b]{0.48\textwidth}
    \centering
    \includegraphics[width=\textwidth]{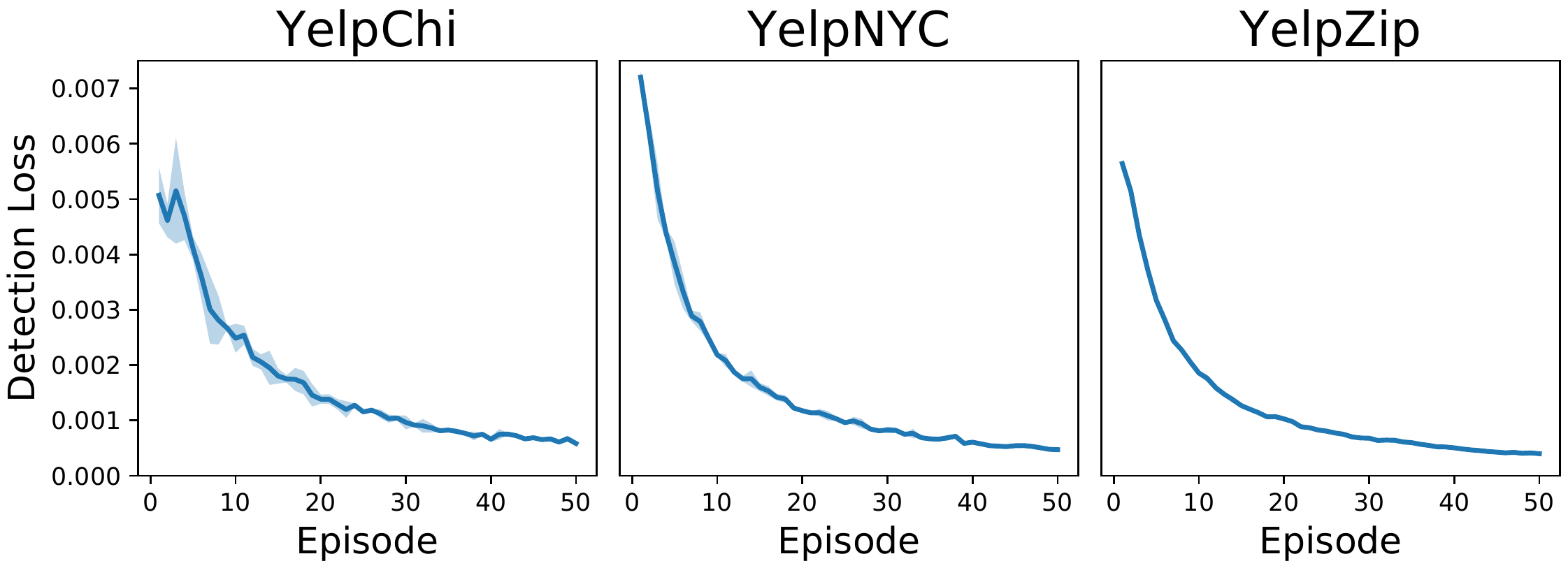}
    \caption{\footnotesize Detector loss ($\mathcal{L}(\mathsf{q})$)}
    \label{fig:detector_loss}
    \end{subfigure}
    \caption{The training process of Nash-Detect with 50 episodes.}
    \label{fig:game_outcome}
\end{figure*}

In the training phase, we run Nash-Detect for 50 episodes on three datasets, and the results are shown in Figure~\ref{fig:game_outcome}.

\vspace{.05in}

\noindent \textbf{Spamming strategy updating.}
Figure~\ref{fig:attack_update} shows the updates that the spamming strategy mixture parameters (\textsf{p}) undergo.
We can see that Nash-Detect successfully identifies reviews posted by new accounts using \texttt{Singleton} as insignificant to the practical effect.
A similar observation is shown in Table~\ref{tab:worst_case} where the \texttt{Singleton} attack's practical effect is much smaller than the other spamming strategies against all detectors.
This observation is consistent with the practical influence of non-elite accounts discussed in Section~\ref{sec:practical_metric}.
In Figure~\ref{fig:attack_update}, the fluctuations in the relative importance ($p_k$ values) of the remaining four spamming strategies indicate that it is hard to find a single useful spamming strategy to evade the defense strategy, which is also under training and evolving.

\vspace{.05in}
\noindent \textbf{Detector importance updating.}
By juxtaposing Figures~\ref{fig:attack_update} and~\ref{fig:detector_update}, we can see that no matter how the spamming strategy evolves, the detectors' weights ($\mathsf{q}$) move smoothly toward the optimal configuration.
Meanwhile, the loss of detectors (Eq. (\ref{eq:detector_loss})) converges as training episodes increases (Figure~\ref{fig:detector_loss}).
Note that in Figure~\ref{fig:detector_update}, though the relative importances of GANG and Fraudar ($\mathsf{q}$) differ across datasets,
the two detectors both perform better than other detectors under the worst cases individual attack under each dataset.
This means that the Nash-Detect converges to the optimal configuration.

\vspace{.05in}
\noindent \textbf{Practical effect.}
According to Figures~\ref{fig:detector_update} and~\ref{fig:practical_effect},
the learned detector importance ($\mathsf{q}$) do not use a single detector but with a stable distribution that guarantees the worst-case performance is better than the worst-case when using a single detector (dash lines at the top).
By comparing Figure~\ref{fig:evaded_reviews} and~\ref{fig:total_return}, we see that the number of evaded spams are not positively correlated with the total rewards of the spamming attacks on the target products.
This observation further confirms that the promotions mainly come from the small number of elite accounts that evaded the detection,
even with many singleton spams detected. 
Similar observations can be seen from Figures~\ref{fig:evaded_reviews} and~\ref{fig:practical_effect},
as the practical effect will be backpropagated to individual products as their rewards.
The bumping of the number of evaded reviews on the YelpNYC dataset is due to the suspicious scores of posted fake reviews are around the filtering threshold.  

\vspace{.05in}
\noindent \textbf{Interpretability}.
Since the fake reviews posted by elite accounts contribute much more than other reviews to practical effect (Eq. (\ref{eq:practical_metric})), Nash-Detect favors detectors that can spot more elite accounts.
From the graph perspective,
the elite accounts and target products are likely to form dense blocks when fake reviews are added between the accounts and the products.
Since Fraudar is designed to spot dense blocks, more elite spams should be detected and Nash-Detect successfully assigns higher importances to Fraudar, as shown in Figure~\ref{fig:detector_update}.
Contrarily, fBox is designed to detect ``small'' attacks and can't handle elite spams.
Since singleton spams are insignificant to the practical influence, Nash-Detect assigns lower importance to fBox.

\subsection{Performance in Deployment}
\begin{figure*}
    \centering
    \includegraphics[width=\textwidth]{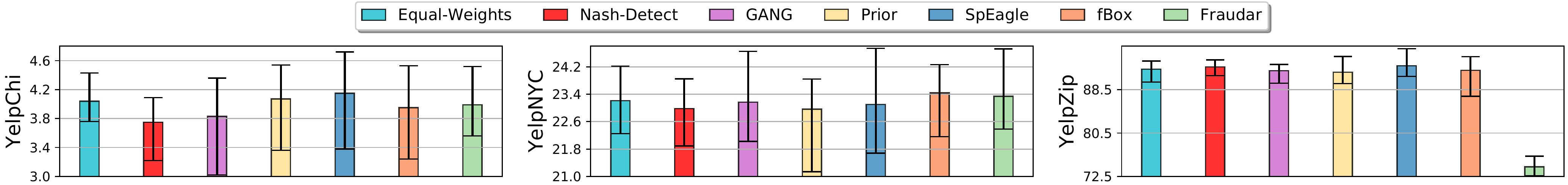}
    \caption{The practical effect of Nash-Detect and baselines during testing.}
    \label{fig:testing}
\end{figure*}

In the testing phase,
we select new sets of controlled accounts following the same settings in the training phase but keep the same set of target products.
For each target product, we sample base spamming strategies uniformly to test the performance of various detectors in suppressing practical spamming effects when deployed. 
We repeat the sampling for 10 times.
In Figure~\ref{fig:testing} we show the means and standard deviations of the practical effects when the detectors are deployed.
Equal-Weights is a baseline where all base detectors have the same importance,
and Nash-Detect represents the optimal detector configuration ($\mathsf{q}^{\ast}$) found during the training phase.
From Figure~\ref{fig:testing},
we can see that Nash-Detect (in red bars) has lower mean practical effects, compared to other detectors on YelpChi and YelpNYC.
On YelpZip, the performance of Nash-Detect is not as good as Fraudar's, since the selected accounts in the testing phase have a higher percentage of elite accounts with more reviews ($>100$).
As a result, Fraudar can easily spot the large dense blocks and has a much lower practical effect on YelpZip.
This suggests that one should incorporate diverse accounts and products during training to guarantee the robustness of the resulting detectors.
More experiment results are presented in the Supplement~\ref{sec:more_sensitivity}.

\section{Related Work}
\label{sec:related_work}
\textbf{Spam detectors.}
Besides text-based detectors~\cite{Mukherjee:2013uk, Rayana2015, wang2017handling, kaghazgaran2019wide}, there are three major types of spam detectors: behavior-based detectors, graph-based detectors, and detectors using both behavioral and graph structural information.
Behavior-based detectors \cite{Rayana2015,Mukherjee:2013uk,mukherjee2013spotting,Xie2012a} calculate review suspicious scores based on different behavior features. Graph-based detectors adopt representation learning approaches, such as SVD~\cite{Shah2014},  node2vec~\cite{kaghazgaran2018combating}, and more recently graph neural networks~\cite{liu2020alleviating}. Some graph-based detectors identify dense blocks~\cite{Hooi2016} or dense flows~\cite{liflowscope} of the spamming accounts and targets from review graph. Examples of detectors using both behavioral and graph structural information include the SpEagle~\cite{Rayana2015}, GANG~\cite{Wang2017}, and GraphConsis~\cite{liu2020alleviating}, etc.

\vspace{.05in}
\noindent \textbf{Marketing research on spam campaigns.}
Online review systems and the associated spam campaigns have drawn much attention from researchers in marketing. \cite{ye2011influence, luca2016reviews, forman2008examining} study how online business ratings influence consumers' purchase decisions. \cite{ye2011influence} shows that a ten percent increase in traveler review ratings will boost online bookings by more than five percent. 
\cite{forman2008examining} suggests that both the reviews and the accounts' social status will contribute to revenue growth.
Inspired by the above works, we propose to incorporate economic incentives and accounts' social status into the spamming practical goal.
In the real world business system, there are other factors like marketing operation status and business category that may affect the business revenue as well.
However, it is infeasible for us to model such a complicated relationship using only the online review data.
Therefore, in this paper, we mainly focus on the relationship between rating and revenue.

\section{Conclusion}
\label{sec:conclusion}
In this work, we propose a practical metric for review systems with product revenue promotion in mind.
We investigate the practical performance of mainstream spam detectors against several goal-oriented spamming attacks in an adversarial setting.
We formulate a game theoretical model and reinforcement learning algorithm to find a robust detection strategy against a diverse set of spamming strategies. 
Empirical evaluations on three large review datasets demonstrate that the proposed algorithm can indeed generate detectors that can effectively tame the practical spamming goal of product revenue manipulation. 
It remains our future work to adopt the proposed model and algorithms to adversarial misinformation attack in other reputation systems, such as rumor detection on social networks.



\begin{acks}
Philip S. Yu and Yingtong Dou are supported in part by NSF under grants III-1526499, III-1763325, III-1909323, CNS-1930941. Sihong Xie is supported by NSF under grants CNS-1931042. 
\end{acks}

\bibliographystyle{ACM-Reference-Format}
\bibliography{sample-base}



\appendix

\section{\textsc{SUPPLEMENT}}
\label{sec:supplement}

\subsection{Implementation Notes}
\label{sec:append_exp}

\subsubsection{Software and Hardware Configurations}
All algorithms are implemented in Python 3.7.3 with standard scientific computation packages. All codes are executed on a 3.50GHz Intel Core i5 Linux desktop with 64GB RAM.

\subsubsection{Dataset Preprocessing Details.}
Table~\ref{tab:statistics} shows the dataset statistics and spamming attack settings in the experiment. We exclude all the labeled spammers and their reviews in the original datasets to avoid biases during the evaluation.
The fake reviews posted by attacks are regarded as positive (spam) instances, and all other reviews are negative (legitimate) instances.

\subsubsection{Spamming Attack Setting}
\label{sec:attack_setting}
Under YelpChi, for non-singleton attacks, the spammer controls the same set of elite accounts with the same targets.
Among the legitimate elite accounts,
we select a fixed set of 100 accounts with minimum suspicious scores obtained by Prior as the controlled elite accounts.
In this way, we can guarantee that an elite account has a small suspicious prior and will always maintain its elite status during the spamming campaign.
We create 450 new accounts for the \texttt{Singleton} attack.
Based on the item features in~\cite{Rayana2015}, we select 30 items with the lowest suspicious scores as the target items.
We post fake five-star reviews in one day for promotion purposes.
The attack settings for YelpNYC and YelpZip are set up similarly with scaled amounts (see Table~\ref{tab:statistics} for details).

\subsubsection{Spam Detector Implementation Details}
\label{sec:detector_details}
\begin{itemize}[leftmargin=*]
\setlength\itemsep{0.05em}
    \item GANG~\cite{Wang2017}: since the original algorithm only outputs the posterior belief of account nodes, we take the average value of account posterior belief and the review prior belief as the review suspicious score. We reproduce the C++ code from authors in Python.

\item SpEagle~\cite{Rayana2015}: we reproduce the original MatLab code in Python. 

\item fBox~\cite{Shah2014}: the original fBox algorithm only outputs a list of detected suspicious nodes. We take the prior belief of reviews posted by detected accounts as review suspicious score and set the suspicious score of other reviews to be zero. We use the Python code provided by authors. Two parameters of the model are $\tau=20\%, k=50$. $\tau$ is optimized from a search space $[1\%, 5\%, 10\%, 20\%, 50\%, 99\%]$ and $k$ is the value from the paper.

\item Fraudar~\cite{Hooi2016}: we use the Python source code provided by authors.
Fraudar is running under the $detect\_blocks$ mode,
where every reviews will be assigned to a dense block,
the output of Fraudar are dense blocks and their density scores.
Similar to fBox, Fraudar only outputs the density score of accounts.
We take the density score of an account as the suspicious score of its reviews.

\item Prior: we implement the algorithm in Python. Review suspicious score is computed via the aggregation function introduced in~\cite{Rayana2015}:
\begin{equation}
    S_{u} = 1- \sqrt{\frac{\sum_{l=1}^{H}h\left(x_{u}^{2}\right)}{H}},
    \label{eq:spamscore}
    \end{equation}
where $\left\{x_{1}, \dots, x_{Hu}\right\}$ is a set of $H$ behavioral feature values of account $u$. The suspicious score of $u$ is aggregated through a feature transformation function $h$ in~\cite{Rayana2015}.
\end{itemize}

For SpEagle and GANG, we use the same parameter setting reported in their papers, and our Python codes have similar detection performances comparing to the original codes.

\subsubsection{Spamming Strategy Implementation Details}
\label{attack_details}


\begin{itemize}[leftmargin=*]
    \setlength\itemsep{0.1in}
    
    \item \texttt{IncBP.} \texttt{IncBP} estimates the suspiciousness of controlled elite accounts using Linearized Belief Propagation (LinBP) proposed by~\cite{Wang2017}.
    LinBP is an improved loopy belief propagation~\cite{murphy1999loopy} algorithm on Markov Random Field (MRF) with less time complexity.
    In this work, the MRF is composed of account nodes and product nodes, the reviews between them are edges.
    For each target, \texttt{IncBP} runs LinBP first and selects the accounts with minimum suspicious scores to post fake reviews.
    Then it updates the entire graph with injected reviews, reruns LinBP, and attacks the next target iteratively.

    \item \texttt{IncDS.} For each target, \texttt{IncDS} first uses all controlled elite accounts to post fake reviews to the target on an auxiliary graph;
    it then calculates each controlled account's density on the auxiliary graph.
    The density of an account $u$ is:
    \begin{equation}
    d(u) = \frac{\left|\mathcal{R}(u)\right|}{\sum_{r_{uv}\in\mathcal{R}}\left | \mathcal{R}(v)\right |}.
    \label{eq:density}
    \end{equation}
    It is the division between the degree of $u$ and the degree of all products reviewed by $u$.
    \texttt{IncDS} selects the accounts with the minimum densities to execute the real fake review injection for the current target.
    Such a mechanism guarantees the spammer avoiding formulating dense blocks around the controlled accounts during the spamming attack.
    
    \item \texttt{IncPR.} \texttt{IncPR} does not rely on graph information of the review system.
    It first calculates the suspicious score of each controlled elite account using Prior. 
    Then it selects the accounts with the minimum prior suspiciousness to add fake reviews.
    
\end{itemize}

\begin{table}[t]
\small
\caption{Dataset statistics and attack settings}
\centering
\resizebox{0.9\linewidth}{!}{%
\begin{tabular}{|l|c|c|c|c|c|c|}  
 \hline
\multirow{3}{*}[0.5em]{\textbf{Dataset}} & \multicolumn{3}{c|}{\textbf{Dataset Statistics}} & \multicolumn{3}{c|}{\textbf{Attack Settings}} \\
\cline{2-7}
 & \textbf{$|\mathcal{U}|$} & \textbf{$|\mathcal{V}|$} & \textbf{$|\mathcal{R}|$} & $|\mathcal{U}_{E}|$ & $|\mathcal{V}_{T}|$ &  $|\mathcal{R}(\mathsf{p})|$ \\ 
\hline
YelpChi & 38063 & 201  & 67395   & 100   &  30  &  450 \\
 
YelpNYC & 160225 & 923  & 359052   & 400   & 120  & 1800  \\
 
YelpZip & 260277 & 5044  & 608598   & 700   & 600 & 9000 \\
\hline
\end{tabular}}
\label{tab:statistics}
\end{table} 


\subsection{Sensitivity Study}
\label{sec:more_sensitivity}

\begin{figure*}
\centering
    \begin{subfigure}[b]{0.33\textwidth}
    \centering
    \includegraphics[width=\textwidth]{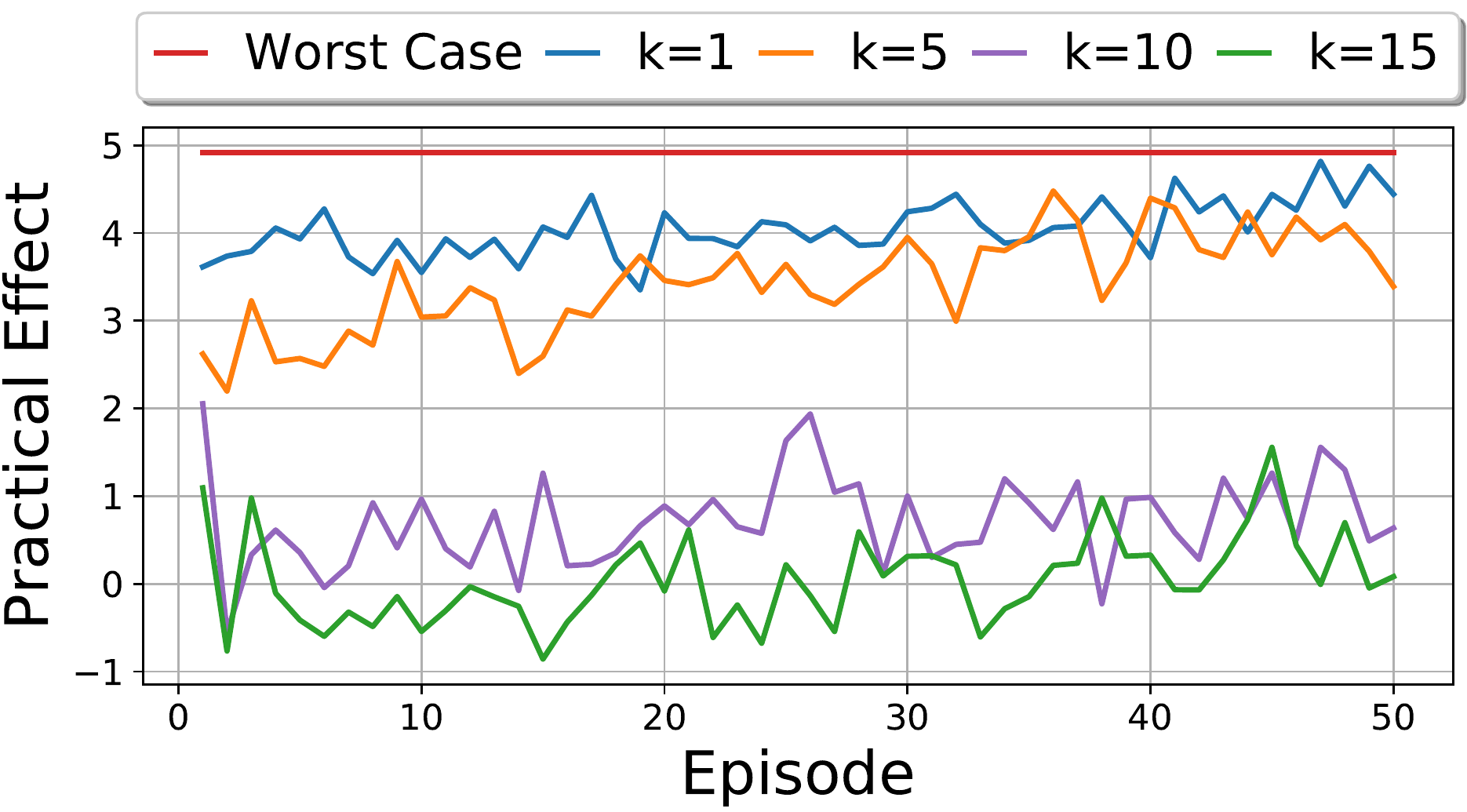}
    \caption{\footnotesize Top $k\%$ sensitivity.}
    \label{fig:top_k}
    \end{subfigure}
    \hfill
    \begin{subfigure}[b]{0.33\textwidth}
    \centering
    \includegraphics[width=\textwidth]{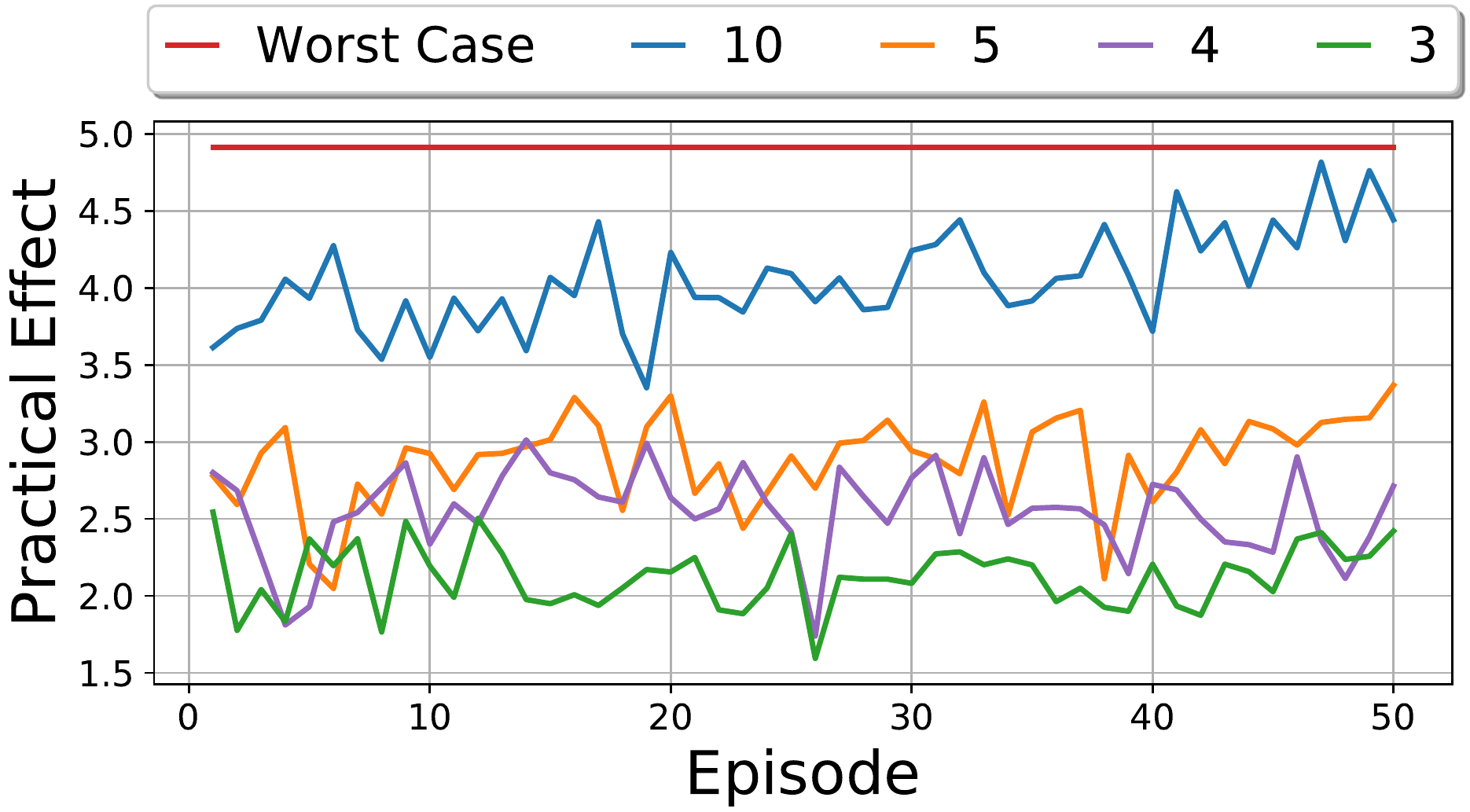}
    \caption{\footnotesize Elite threshold.}
    \label{fig:elite_threshold}
    \end{subfigure}
    \hfill
    \begin{subfigure}[b]{0.33\textwidth}
    \centering
    \includegraphics[width=\textwidth]{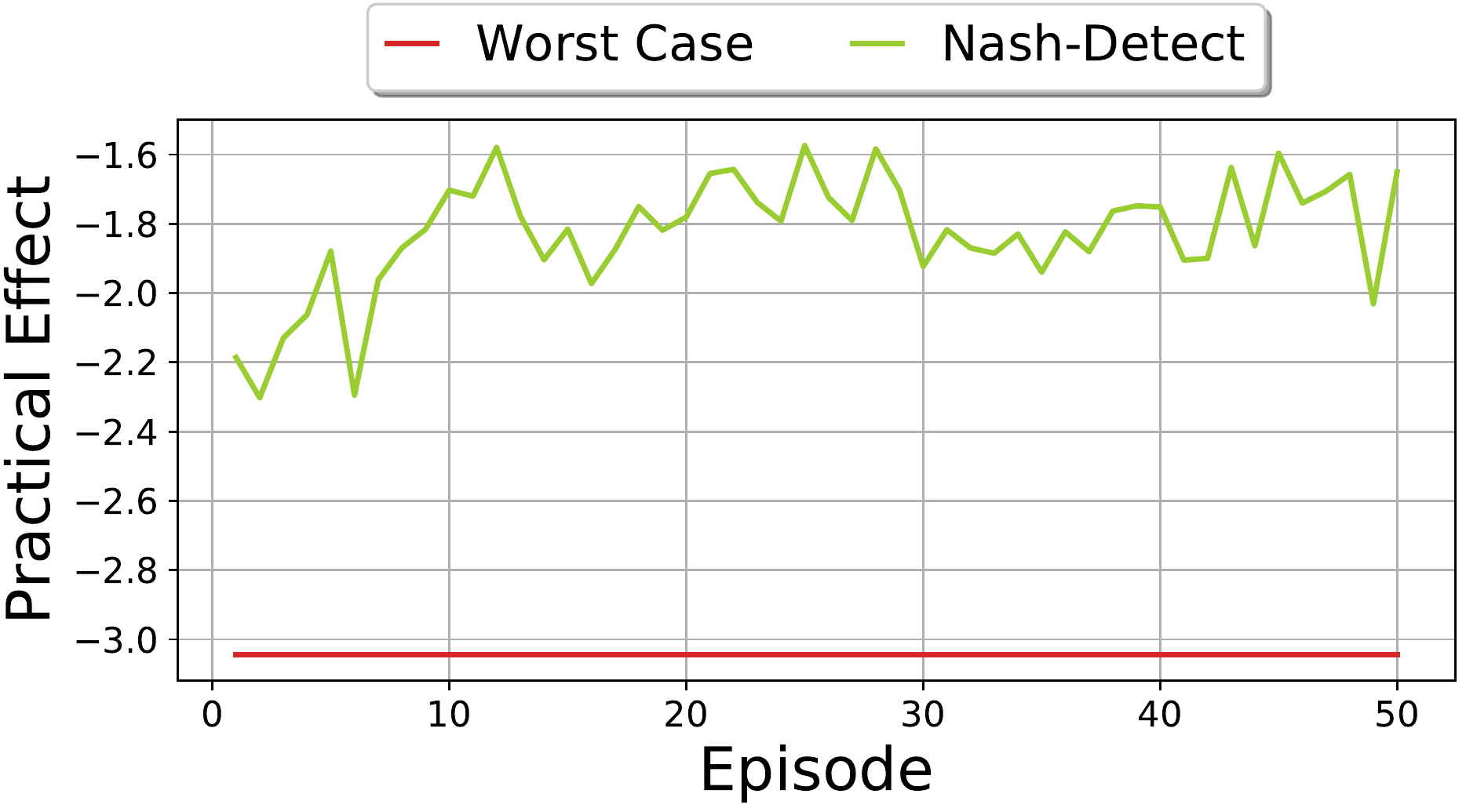}
    \caption{\footnotesize Demoting attack.}
    \label{fig:demoting}
    \end{subfigure}
     \caption{\footnotesize The practical effect of (a) different top k\% thresholds; (b) different elite account selection thresholds; (c) demoting attacks under Yelpchi.}
\end{figure*}

\noindent \textbf{Top $k$.}
Figure~\ref{fig:top_k} shows the practical effect change under different detection filtering thresholds.
As the filtering threshold increases,
the attacks become harder to reach their practical goals;
thus the practical effects of them become lower. 
Under Top $10\%$ and Top $15\%$, the practical effect will come below zero at some episodes.
From Eq. (\ref{eq:revenue}),
the RI is of a product $v$ is obtained by its average rating minus the average rating of all products.
When we increase the filtering threshold,
more fake reviews of the target products will be removed.
Meanwhile, the remaining reviews will increase the average rating of all products ($g(\mathcal{R})$).
It will make the $g(\mathcal{R}(v))$ less than $g(\mathcal{R})$ for some target product $v$.
Consequently, the practical effect will become zero at some episodes.

\vspace{.1in}
\noindent \textbf{Elite account threshold.}
In Section~\ref{sec:exp_settings}, we set $\#review\geq 10$ as the elite account selection threshold.
Figure~\ref{fig:elite_threshold} presents the experiment results on more thresholds,
where 5, 4, and 3 select top $5.7\%, 8.5\%, 14\%$ accounts with most reviews.
We can see that attack could reach the best practical effect with \textit{ten} as the threshold.
The reason is that,
under a greater elite account selection threshold, the spammer controls a greater proportion of elite accounts among all accounts.
Thus, those controlled elite accounts have more significant impact on the practical effect.

\vspace{.1in}
\noindent \textbf{Demotion attack.}
Previous experiments focus on the promotion spamming attack because it is more prevalent. 
According to~\cite{luca2016reviews}, the demoting spamming attacks usually appear between competitors.
We change our attacks to demoting attacks by modifying the rating of posted reviews from five stars to one star.
Figure~\ref{fig:demoting} shows the practical effect of demoting attacks on YelpChi.
Note that the practical effect (Eq. (\ref{eq:practical_metric})) is negative for the demoting spam campaign.
Without changing the initial configuration,
the proposed algorithm can still minimize the negative influence of demoting attacks,
which is always less than the worst case. 

\begin{figure}
\centering
     \includegraphics[width=0.48\textwidth]{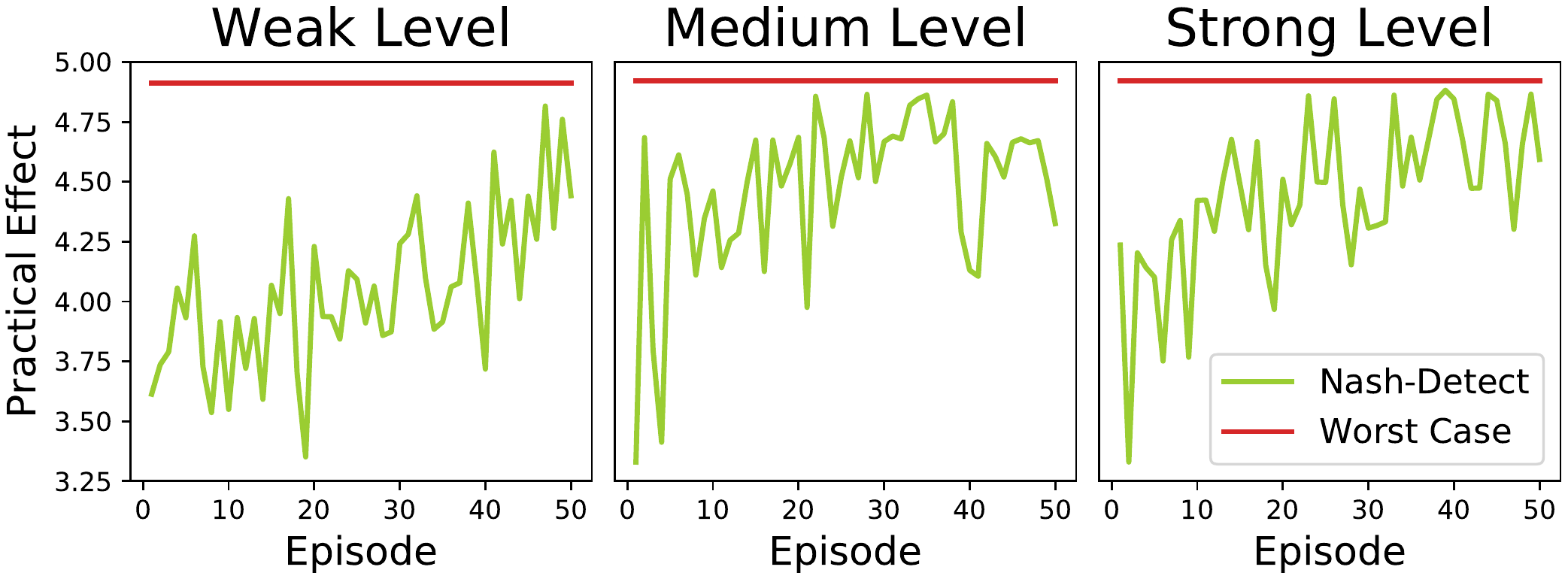}
     \caption{\footnotesize The game outcome under different behavior camouflage levels.}
     \label{fig:behavior}
\end{figure}

\vspace{.1in}
\noindent \textbf{Spammer's behavior.}
In real-world spamming campaigns, since most of the merchants soliciting fake reviews aim to meet the spamming goal as fast as possible~\cite{Xie2012a, Mukherjee:2013uk}, spammers must finish fake review injection tasks within a limited period.
However, fruitful features and algorithms have been proposed to capture the burstiness (or spikes) of review activities to spot spam campaigns~\cite{zheng2017smoke, Rayana2015}.
Correspondingly, strategic spammers may carefully adjust the posting frequency to avoid traffic spikes~\cite{kaghazgaran2019tomcat}.
Besides that, spammers could smooth out their reviews with moderate ratings (i.e., two or four stars out of five) to evade the detection~\cite{rahman2019art}.

To study the sensitivity of our algorithm on different behaviors of the spammer under promotion attack,
we take the same setting in previous experiments, where all 5-star fake reviews are posted in one day, as a \textit{Weak Level Attack}.
Then we propose the \textit{Medium Level Attack}: fake reviews are posted in five days, and their ratings are randomly sampled from four to five stars.
The \textit{Strong Level Attack} is where fake reviews are posted within fifteen days.
Review ratings are selected in the same way as the \textit{Medium Level Attack}.
Figure~\ref{fig:behavior} shows that no matter how spammer switch adjust their behavior, Nash-Detect can make the practical effect always less than the worst case.

\begin{figure}
\centering
    \begin{subfigure}[b]{0.49\textwidth}
    \centering
    \includegraphics[width=\textwidth]{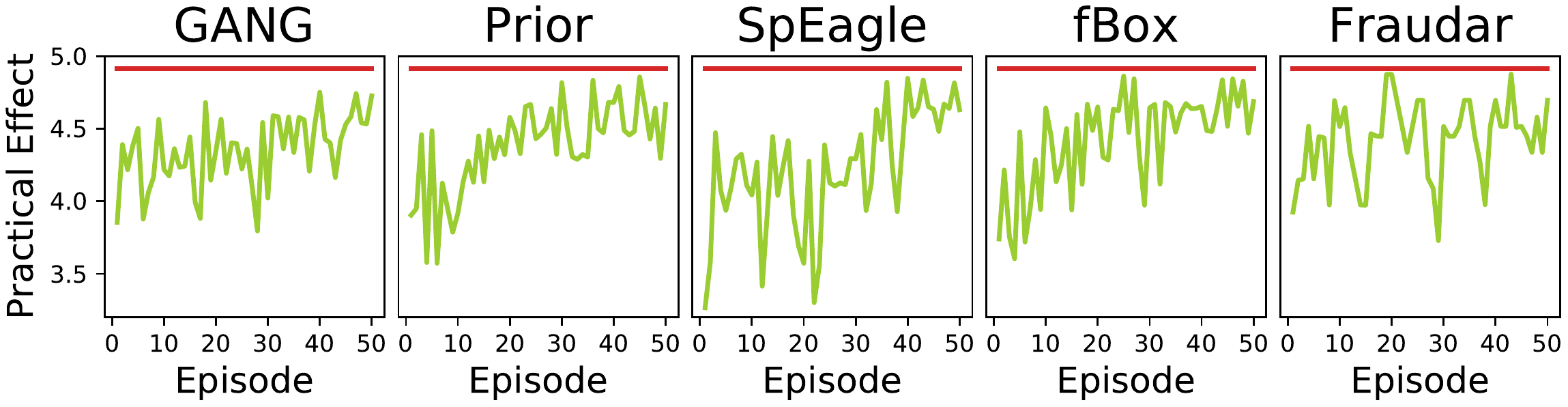}
    \end{subfigure}
    \bigskip
    \begin{subfigure}[b]{0.49\textwidth}
    \centering
    \includegraphics[width=\textwidth]{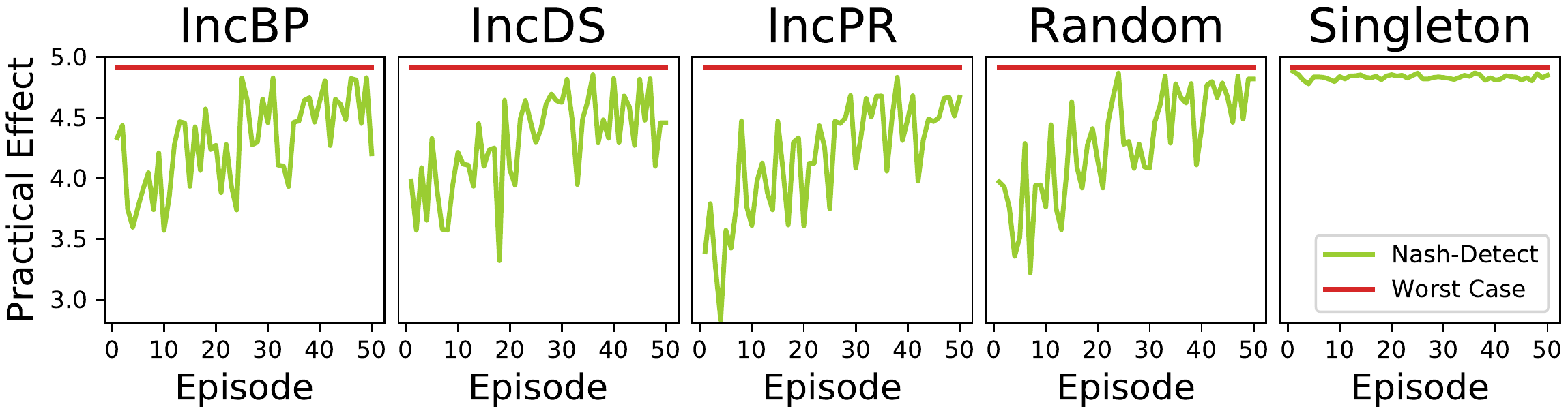}
    \end{subfigure}
     \caption{\footnotesize The practical effect of Nash-Detect vs. the worst case of games with less detectors and attacks under YelpChi. The title of each sub-figure represents the removed detector/attack.}
     \label{fig:miss_one}
\end{figure}

\vspace{.1in}
\noindent \textbf{Sensitivity on the base detector and attack amounts.}
Each sub-figure in Figure~\ref{fig:miss_one} shows the practical effect of the game when one detector/attack is removed (figure title indicates the removed one).
We could see that our algorithm still performs better than the worst case with any four attacks/detectors.
The second row of Figure~\ref{fig:miss_one} shows that the first four attacks with controlled elite accounts have similar performance,
and the \texttt{Singleton}'s plot indicates that the spammer could get a better practical effect only using elite accounts. 

\end{document}